\documentclass{aa}  

\usepackage{graphicx}
\usepackage{txfonts}
\defcitealias{GEMSI}{GEMS I}
\defcitealias{GEMSII}{GEMS II}
\defcitealias{GEMSVII}{GEMS VII}
\defcitealias{GEMSIV}{GEMS IV}
\usepackage[normalem]{ulem}

\begin{document}
\newcommand{\trans}[2]{$(#1 \rightarrow #2)$}

   \title{Gas phase Elemental abundances in Molecular cloudS (GEMS)}

   \subtitle{X: Observational effects of turbulence on the chemistry of molecular clouds}

   \author{L. Beitia-Antero
          \inst{1,4}
          \and
          A. Fuente\inst{2} 
          \and
          D. Navarro-Almaida\inst{3}
          \and
          A.I. G\'omez de Castro\inst{4,5}
          \and
          V. Wakelam\inst{6}
          \and
          P. Caselli\inst{7}
          \and
          R. Le Gal\inst{8,9}
          \and
          G. Esplugues\inst{10}
          \and
          P. Rivière-Marichalar\inst{10}
          \and
          S. Spezzano\inst{7}
          \and
          J. E. Pineda\inst{7}
          \and
          M. Rodríguez-Baras\inst{10}
          \and
          A. Canet\inst{4,5}
          \and
          R. Martín-Doménech\inst{2}
          \and
          O. Roncero\inst{11}}

   \institute{Departamento de Estadística e Investigación Operativa, Facultad de Ciencias Matemáticas, Universidad Complutense de Madrid\\
              \email{lbeitia@ucm.es}
         \and
            Centro de Astrobiolog\'{\i}a (CAB), CSIC-INTA, Ctra. de Ajalvir Km. 4, Torrej\'on de Ardoz, 28850 Madrid, Spain 
         \and
            Universit\'{e} Paris-Saclay, Universit\'e Paris Cit\'{e}, CEA, CNRS, AIM, F-91191 Gif-sur-Yvette, France
         \and
            Joint Center for Ultraviolet Astronomy, Universidad Complutense de Madrid, Avda Puerta de Hierro s/n, 28040, Madrid, Spain
         \and
            Sección Departamental, Departamento de Física de la Tierra y Astrofísica, Facultad de Ciencias Matemáticas, Universidad Complutense de Madrid, Plaza de Ciencias 3, 28040 Madrid, Spain
         \and
            Laboratoire d’astrophysique de Bordeaux, Univ. Bordeaux, CNRS, B18N, allée Geoffroy Saint-Hilaire, 33615 Pessac, France
         \and
            Centre for Astrochemical Studies, Max-Planck-Institute for Extraterrestrial Physics, Giessenbachstrasse 1, 85748 Garching,
Germany
         \and
            Institut de Plan\'etologie et d'Astrophysique de Grenoble (IPAG), Universit\'e Grenoble Alpes, CNRS, F-38000 Grenoble, France
         \and
   Institut de Radioastronomie Millim\'etrique (IRAM), 300 Rue de la Piscine, F-38406 Saint-Martin d'H\`{e}res, France
        \and
        Observatorio Astron\'omico Nacional (OAN, IGN), Alfonso XII, 3, 28014, Madrid, Spain
         \and
         Instituto de Física Fundamental (CSIC), Calle Serrano 121–123, 28006 Madrid, Spain\\}

   \date{Received XX YY, ZZ; accepted XX YY, ZZ}

 
  \abstract
   { We explore the chemistry of the most abundant C-, O-, S-, and N-bearing species in molecular clouds, in the context of the IRAM 30 m Large Programme Gas phase Elemental abundances in Molecular Clouds (GEMS). Thus far, we have studied the impact of the variations in the temperature, density, cosmic-ray ionisation rate, and incident UV field in a set of abundant molecular species. In addition, the observed molecular abundances might be 
   affected by turbulence which needs to be accounted for
   in order to have a more accurate description of the chemistry of interstellar filaments.}
   {In this work, we aim to 
   assess the limitations introduced in the observational
   works when a uniform density is assumed along the
   line of sight for fitting the observations, developing
   a very simple numerical model of a turbulent box.
   We searched for any
   observational imprints that might provide useful information on the 
   turbulent state of the cloud based on kinematical or chemical tracers.}
   {We performed a magnetohydrodynamical (MHD) simulation
   in order to reproduce the turbulent steady state of a 
   turbulent box with properties typical
   of a
   molecular filament
   before collapse. We
   post-processed the results of the MHD simulation with a chemical
   code to predict molecular abundances, and then post-processed this
   cube with a radiative transfer code to create synthetic emission maps
   for a series of rotational transitions observed during the GEMS project.}
   {From the kinematical point of view, we find that the relative alignment
   between the observer and the mean magnetic field direction affect
   the observed line profiles, obtaining larger line widths for the case
   when the line of sight is perpendicular to the magnetic field. These differences
   might be detectable even after convolution with the IRAM 30 m efficiency
   for a nearby molecular cloud. From the chemical point of view, we find
   that turbulence produces variations for the predicted abundances, but they
   are more or less critical depending on the chosen transition and the
   chemical age. When compared to real observations, the results from
   the turbulent simulation provides a better fit than when assuming a
   uniform gas distribution along the line of sight.}
   {In the view of our results, we conclude that taking into account turbulence
   when fitting observations might significantly improve the agreement with model predictions. This is especially important
   for sulfur bearing species which are very sensitive to the variations of density produced by turbulence at early times (0.1 Myr). 
   The abundance of CO is also quite sensitive to turbulence when considering the evolution beyond a few 0.1 Myr.}

   \titlerunning{Observational effects of turbulence}
   \maketitle
%

\section{Introduction}

Molecular clouds are highly turbulent \citep{2012A&ARv..20...55H,2004ApJ...615L..45H} 
and
although the nature of this turbulence is not fully understood,
it seems to be
a mixture of solenoidal (divergence-free) and compressive
(curl-free) modes, where the contribution of each mode depends on
the local conditions \citep{2010A&A...512A..81F}: regions with
a low star-formation rate seem to be dominated by solenoidal modes,
while massive star-forming complexes with strong feedback present
a non-negligible contribution of compressive modes.
Despite our limited knowledge on the properties of turbulence, it has
been put forward many times as a likely explanation for unexpected
observational results.
For instance, \citet{2018ApJ...862....8T}
found structures in molecular gas that were thought to be caused by
supersonic turbulence.
One of the most puzzling findings was the
high abundance of some molecular species in the diffuse phase of the interstellar
medium (ISM, \citealt{1995ApJS...99..107C, 1996A&A...307..237L, 2012A&A...542L...6N}), 
where according to the predictions of chemical models they should not be
able to survive. In order to determine why this occurred, some theories have been
developed based on the intermittency of turbulence \citep{2014A&A...570A..27G} that aim
at linking the properties of turbulence with the observations. More recently,
the advent of modern computation has made it possible to follow the chemical evolution
of the diffuse phase in 3D magnetohydrodynamical (MHD) simulations, either simultaneously to the
MHD evolution
\citep{2018MNRAS.481.4277F, 2021MNRAS.500.3290M}
 or in a post-processing stage \citep{2015MNRAS.453.2747M, 2017ApJ...850...23B, 2019ApJ...885..109B,
2022arXiv220910196G}. \par

The simultaneous study of the chemical and dynamical evolution of molecular clouds
has a huge potential that only recently has been able to be exploited thanks to
the widespread access to supercomputers. Generally speaking, there are three alternatives
for coupling chemistry and MHD simulations. The first one relies on the implementation
of a reduced chemical network where only a set of species is followed
\citep{2017MNRAS.472.4797S, 2022MNRAS.512..348K}. Depending on the problem to be
studied, the number of species varies and specific chemical networks
can be selected for particular cases, for instance, for the study of
deuterium fractionation in turbulent cloud cores \citep{2017MNRAS.469.2602K, 2021MNRAS.502.1104H},
for the chemistry in
 star-forming filaments \citep{2016MNRAS.459L..11S}, or for assessing
 the reliability of determining filament masses and widths from CO
 emission maps \citep{2017MNRAS.467.4467S}. A second approach
consists of post-processing a numerical simulation with a chemistry code, which is
less expensive in terms of computation time.
This technique has proved to be useful for tasks such as the calibration
of the HCN-star formation correlation \citep{2018MNRAS.479.1702O}
or the analysis of the interface of colliding molecular clouds
\citep{2017ApJ...850...23B}. Finally, the third approach that is also computationally expensive
but allows the chemo-dynamical evolution of the gas to be solved
relies on evolving a set of tracer particles that store the
local properties of the gas at fixed times that can be later post-processed with
chemical codes (\citealt{2013ApJ...775...44H, 2016ApJ...822...12H, 2020A&A...643A.108C, 2021MNRAS.505.3442F, 2024A&A...685A.112N}). \par

It is extremely complicated to establish a direct link between
observations and predictions from numerical models; a recent example is the
work by  \citet{2017MNRAS.471.1506K}, where they introduce several statistical
tracers that can be used for the comparison of simulations and observations; however, when
applied to real datasets they obtained nonconclusive results.
In this article, we present a detailed analysis of the observational effects
of turbulence as predicted by a simple MHD model of a molecular cloud
that we post-processed with a chemical
model plus a radiative
transfer code to analyse the predicted chemical variations
of the main set of molecules observed
during the IRAM 30 m Large Programme
Gas phase Elemental abundances in Molecular cloudS (GEMS) \citep{GEMSI, GEMSII, GEMSIV,
Bulut2021, Spezzano2022, Esplugues2022, GEMSVII}. First, in Sec. \ref{sec:mhd_models} we present the 
setup of the MHD simulation. Then, in Sec. \ref{sec:chemistry}
we give the details on the chemistry and radiative
transfer codes used for post-processing
the simulations in order to derive chemical abundances. The results are presented
in the following sections focussing on: (i) the kinematic tracers in
Sec. \ref{sec:line_profiles}, with a special focus on the effects on the
spectral line profiles; and
(ii) the effects on the estimated chemical abundances (Sec. \ref{sec:abundances}).
In Sec. \ref{sec:discussion} we provide
a comparison between our predictions and a subset of the full GEMS
observational sample, 
and in Sec. \ref{sec:limitations} we discuss the limitations
of the paper.
Finally, in Sec.
\ref{sec:conclusions} we provide the conclusions of our work.



\section{MHD model}\label{sec:mhd_models}

Ideally, we would like to simulate the evolution 
of a molecular cloud from the moment of its formation 
up to the point where star formation might occur, just
before collapse, taking into account all the physics
and chemistry involved. However, this kind of simulation
is highly expensive in terms of computational time and
requires implementing a reduced chemical network that 
reproduces properly the abundances of the main sulfur
reservoirs, which does not exist at this point. Therefore,
the aim of this paper
was to develop a very simple numerical model that allowed us
to assess the extent of the influence of turbulence on real 
single-dish
observations. In particular, our goal was to analyse the results of these
simulations in the context of the GEMS project.
 Therefore, although numerical models withouth
 chemistry can be generalised and do not have an inherent
 physical scale unless some microphysiscs is included, we
wanted to choose the parameters such as they could be
representative of the interior of a low-mass star formation
filament such as the Taurus molecular cloud, which we have
extensively studied in previous papers of the GEMS series.
If we were to model a 
molecular
filament of width 0.5 pc, according to the following equation
\citep{2022arXiv220309562H}: 

\begin{equation}
\langle n \rangle \simeq (1.2 \pm 0.3) \times 10^{4} \bigg( \frac{L}{\rm pc} \bigg)^{-1.1 \pm 0.1} {\rm cm}^{-3}
\end{equation}

corresponds to a particle density of
$n = 2.57 \times 10^{4}$ cm$^{-3}$, and a magnetic field strength 
of 92.60 $\mu$G ($B = 10~\mu$G$~(n/300~{\rm cm}^{-3})^{0.5}$, \citealt{2015MNRAS.451.4384T}).
In principle, modelling the MHD evolution of
such filament is trivial, but in a post-processing stage
we are aiming to recover molecular line emission maps
for the transitions observed in GEMS. Considering that
the nominal resolution of IRAM 30 m telescope
for most of the transitions observed
in GEMS is around 25\arcsec, for the case of a nearby cloud
at the distance of 
Taurus ($\sim$ 140 pc, \citealt{2007ApJ...671..546L}) roughly corresponds to
0.01 pc. 
For the study of observational dilution effects of small scale turbulent patches
we consider that in order to derive robust statistical estimates, we need to average
at least 20 cells per beam so that we reproduce
the same `loss' of resolution than observations, and therefore the optimal
resolution for the simulation would be 1024$^{3}$ cells, which is 
highly expensive in terms of computer time for both the MHD simulation and the post-processing stage with radiative transfer. Instead, we opted for performing small
scale simulations (box size of 0.05 pc and numerical resolution of
256$^{3}$) representative of a portion
of the full filament.
Depending on the observed frequency,
this box contains 2-3 beams of the IRAM 30 m
telescope assuming the distance of Taurus. This resolution, although
quite low for standard MHD turbulence models, is enough to
account for the observational effects since we will be
averaging the density values contained in a beam. This, of course, will bias
our study to large-scale density fluctuations as discussed in Sec. \ref{sec:limitations}.\par 

For the MHD model, we adopt the turbulent-box approximation.
This framework is very common for studying turbulent mixing in
the ISM 
\citep{2017ApJ...843...92B,2021MNRAS.500.3290M, 2023A&A...669A..74G}
and assumes a medium initially at rest in
which turbulence is injected over time via a forcing source.
In this paper, we use the Athena++\footnote{\url{https://github.com/PrincetonUniversity/athena}}
\citep{2020ApJS..249....4S} MHD code, with 
a second-order Van-Leer time integrator,
the Harten-Lax-van Leer-Discontinuities (HLLD) Riemann solver and second-order
spatial reconstruction (Picewise-Linear-Mesh, PLM). In the
current public version of Athena++, a turbulence driver
based on the Ornstein-Uhlenbeck process \citep{2012ApJ...758...78L}
is implemented; the full details of this implementation can be
found in \citet{2012ApJ...758...78L}, but we will provide here
a simple explanation for the general reader. In a nutshell,
the turbulence driver injects velocity `kicks' through
a velocity perturbation which amplitude is chosen randomly
in the Fourier space normalised by a decreasing power-law
in $k$, ensuring that the majority of the energy is injected
in the largest scale, which is $L/2$, where $L$ is the domain
length. This random velocity field which is injected in the domain
allows for a combination of solenoidal and compressive modes, and
the net energy input can be also controlled so that the
desired turbulent Mach number is achieved.

In this initial exploratory work, we assumed an isothermal medium that evolves
according to the equations of ideal MHD.
Although this approach is rather simplistic 
because we are ignoring the ion-neutral drift and assuming no heating or cooling
effects, this model can be regarded as a starting point
for estimating the effects of turbulence in the predicted molecular
abundances of a molecular filament; in a future work, we will
refine it including more physics as well as active chemistry. Besides, as pointed out by \citet{2017ApJ...843...92B}
for moderate mach numbers ($\mathcal{M} \lesssim 5 $) the gas distribution
for isothermal and non-isothermal simulations is similar, and since 
observations of non-fragmenting molecular filaments are compatible with
a roughly transonic state
\citep{2022arXiv220309562H}, we can safely adopt these assumptions as long as
we keep in mind that the estimates will be somewhat limited;
we discuss the current limitations of our model in Section \ref{sec:limitations}. \par

As we have already mentioned, the MHD simulations are 
scale-free and the main parameters could be chosen based on the
required Mach number, the plasma $\beta$ parameter, and some
characteristic length. However, in order to provide an easier
guide for observers and to ensure that the environment is as 
representative for Taurus as possible,
we set up the numerical parameters
based on physical quantities for gas density, temperature and
box length. We assume that we have an isothermal medium
at $T = 12$ K, $n = 2.57 \times 10^{4}$ cm$^{-3}$
and a box size of $L = 0.05$ pc at a resolution
of 256$^{3}$ with periodic boundary
conditions. We choose a wavenumber for the
turbulent spectrum between $2\pi/L$ and
$4\pi/L$ following a power law $P(k) \propto k^{-2}$, and
assuming a scenario where we have both solenoidal and
compressive modes in the same proportion (mixed turbulence).
This is a reasonable assumption since Taurus is a low-mass 
star formating region, and compressive turbulence is more
characteristic of high-mass star forming regions, but we cannot
discard the presence of some shock waves propagating in the
cloud due to former stellar feedback. In any case, in 
Sec. \ref{sec:limitations} we discuss the differences between
this model of mixed turbulence with other similar ones with different
turbulent mixing modes and magnetic field strengths, which are 
negligible.

\begin{figure}
\centering
\includegraphics[width = \columnwidth]{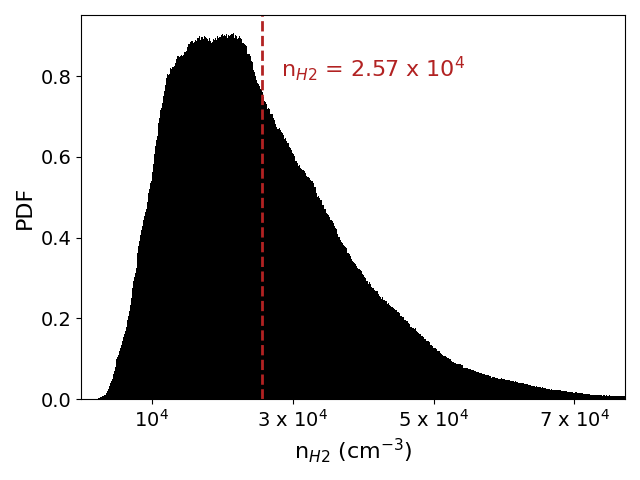}
\includegraphics[width = \columnwidth]{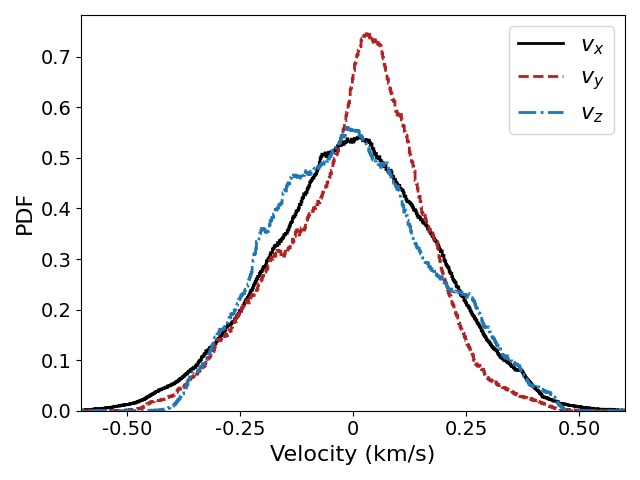}
\caption{Density and velocity probability density functions (PDFs)
at the final stage of the simulation ($\sim 0.36$ Myr).}
\label{fig:simulation_PDFs}
\end{figure}

We start from a uniform medium at rest
with magnetic field parallel to the x-axis and a strength of
$B_{0} = 40~\mu$G which is less than
the roughly $90~\mu$G predicted by \citet{2015MNRAS.451.4384T},
but compatible with observations (see Fig. 5 in 
\citealt{2022arXiv220309562H}). 
In order to reproduce the steady-state of a real cloud,
we continuously drive turbulence until equilibrium is reached
between injection and dissipation rates, which typically occurs
after two crossing times, where we define $t_{cross} = L/2c_{s}\mathcal{M}$ as in \citet{2013MNRAS.436.1245F} and
\citet{2022MNRAS.510.4767L}, and we let the simulation to
evolve for three crossing times, resulting in a 
trans-sonic turbulent
box evolved for roughly 0.36 Myr with velocity dispersions
of the order of 0.23 km/s (see Fig. \ref{fig:simulation_PDFs}), comprised in the range
0.1 - 0.5 km/s typical of Taurus \citep{2018ApJ...864...82D}. The
resulting density and velocity probability
density functions (PDFs) are shown in Fig \ref{fig:simulation_PDFs}.
In the density PDF, the values range from
$0.06 n_{0}$ and $5.95 n_{0}$, where $n_{0} = 2.57 \times 10^{4}$ cm$^{-3}$ 
is the initial molecular hydrogen density. Other important quantities that will
be used later in the paper are the 1.5\% and 98.5\% percentiles, which
correspond to $P_{1.5} = 0.27 n_{0}$ and $P_{98.5} = 2.41 n_{0}$.
We want to note that in our simulation, the mean direction
of the magnetic field barely changes due to turbulence (mean deviation of $1.5^{\rm o}$), providing a
scenario coherent with Taurus 
observations reporting that 
the magnetic field is relatively uniform in scales of pc and roughly perpendicular to the molecular filaments \citep{1997ApJ...487..314M, 2011ApJ...741...21C}. The apparent differences in the velocity PDFs are not physical but a consequence of the moderate numerical resolution.

\section{Chemical models and post-processing}\label{sec:chemistry}

As in previous GEMS works (\citealt{GEMSII}, hereafter \citetalias{GEMSII}, \citetalias{GEMSIV}), for the chemical
modelling we use \textsc{Nautilus}
\citep{2016MNRAS.459.3756R}, a three-phase model that solves the kinetic
equations for gas phase and grain surface/mantle reactions and allows
to compute the time-evolution of chemical species for a given physical
structure. Among the numerous input parameters of the model, there
are few that are critical and mostly determine the results:
gas and dust temperatures, visual extinction ($A_{V}$), particle density (in hydrogen
nuclei, $n_{\rm H}$), the cosmic-ray ionisation rate ($\zeta$), the
incident ultraviolet radiation field ($\chi$), and dust-related properties
such as the dust-to-gas ratio, and the size and composition of dust
particles. For our chemical model we take
values compatible with those derived in
\citet{GEMSVII} (hereafter \citetalias{GEMSVII}) for Taurus: a radiation field of
$\chi = 5$ (where a radiation field of $\chi = 1$ corresponds to the 
interstellar radiation field as parametrised in \citealt{2001ApJS..134..263W}),
a cosmic-ray ionisation rate per H$_{2}$ of $\zeta =  10^{-16}$ s$^{-1}$,
and the initial abundances with respect to H displayed in Table
\ref{tab:initial_abundances}. We also explored
the influence of starting from different initial abundances
based on a prephase model and have found little differences
for the case analysed in this work (see the full discussion
in Appendix \ref{appendix:prephase_chemistry}). We
assume that the gas density for the simulations in Sec. \ref{sec:mhd_models}
corresponds to molecular hydrogen, so we have $n_{\rm H} = 5.14 \times 10^{4}$
cm$^{-3}$ and therefore can assume that gas and dust are thermalised
\citep{2017ApJ...843...63F} with $T = 12$ K. Finally, we set the visual extinction value at $A_{V} = 4$ mag,
a value that is consistent given the column densities obtained
(see Fig. \ref{fig:column_density_maps}) and that is
 valid for comparison with observations with a total
line-of-sight visual extinction up to $A_{V} = 8$ mag.
With respect to the dust properties, we considered
a population of silicate grains with internal density $\rho^{\rm int} = 3.5$ g cm$^{-3}$
and a size of 0.1 $\mu$m. For the time being, we assumed a constant dust-to-gas
ratio of 0.01, although we have strong evidence that a perfect correlation is
not realistic \citep{2016MNRAS.455.3570G, 2021ApJ...908..112B}; we 
defer for
a future work the analysis of the effects of varying dust-to-gas ratio on the
chemistry of molecular clouds. The remaining input parameters of the
chemical network, namely the surface parameters, gas and solid species, 
reactions on solid and gas phase, and activation energies
are taken from the standard chemical network provided together with
the public version of \textsc{Nautilus}\footnote{\url{https://forge.oasu.u-bordeaux.fr/LAB/astrochem-tools/pnautilus}}
which makes use of the KIDA\footnote{\url{https://kida.astrochem-tools.org}} database.
\par

\begin{table}
\centering
\caption{Initial elemental abundances for the chemical model with respect to H.}
\label{tab:initial_abundances}
\begin{tabular}{cc}
\hline \hline
Species & Abundance \\
\hline
He & 0.09 \\
N & $6.2\times 10^{-5}$ \\
O & $2.4 \times 10^{-4}$ \\
H$_{2}$ & 0.5 \\
C$^{+}$ & $1.7 \times 10^{-4}$ \\
S$^{+}$ & $7 \times 10^{-7}$ \\
Si$^{+}$ & $8 \times 10^{-9}$ \\
Fe$^{+}$ & $3 \times 10^{-9}$ \\
Na$^{+}$ & $2 \times 10^{-9}$ \\
Mg$^{+}$ & $7 \times 10^{-9}$ \\
P$^{+}$ & $2 \times 10^{-10}$ \\
Cl$^{+}$ & $1 \times 10^{-9}$ \\
F & $6.68 \times 10^{-9}$ \\
\hline
\end{tabular}
\end{table}

\begin{figure}
\centering
\includegraphics[width = \columnwidth]{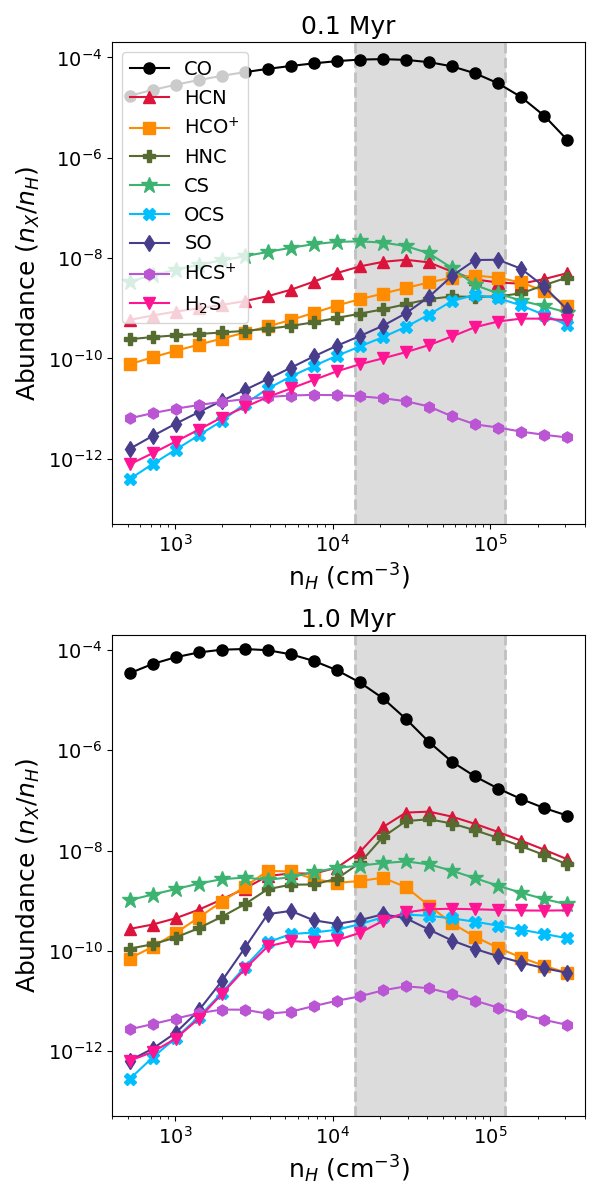}
\caption{Chemical models computed with \textsc{Nautilus} for 0.1 Myr (top) and 1.0 Myr (bottom). 
The densities covered
by the grids are defined so that the minimum and maximum values
reached by the simulations are included. The shaded region in both
plots corresponds to the density range $P_{1.5} \leq \rho \leq P_{98.5}$,
so most of the simulation cells (97\%) are included in this interval.}
\label{fig:chemical_models}
\end{figure}

Deriving the chemical abundances for a given set of physical parameters
with \textsc{Nautilus}
is relatively easy and takes only a few ($\sim 30$) seconds. However,
if we run \textsc{Nautilus} for each of the simulation cells ($256^{3}$),
post-processing a simulation would take almost 6000 h, which is unfeasible.
Instead, we analysed the behaviour of some of the molecules observed
during the GEMS project at the densities of interest for our simulation,
in order to ascertain if some kind of interpolation was possible. In the view
of the results displayed in Fig. \ref{fig:chemical_models}, we concluded that
a chemical network of 20 density nodes (in logarithmic scale)
was enough to derive the chemical abundances
for our simulation, considering a linear interpolation at intermediate values.
Besides, during the development of the GEMS project we also
found that most of the times it was impossible to fit all the observations
with a unique chemical age; in fact, low-extinction lines of sight in
Taurus are better fitted with models of 1 Myr, while higher
visual extinctions are better modelled by 0.1 Myr 
\citepalias{GEMSVII}. In consequence, we also considered
two chemical ages for the post-processing (1 Myr and 0.1 Myr) as they are
bound to produce fundamentally different results in the view of the curves
shown in Fig. \ref{fig:chemical_models}. Of special importance is the large
variance of the predicted CO abundance for the chemical model at 1 Myr, since
it spans more than two orders of magnitude for most of the simulation 
cells (shaded regions). Other molecules typically observed such as HCN, HNC or
HCO$^{+}$ also present a slightly variable behaviour, 
being more pronounced again for the model at 1 Myr. In this study we
also included a series of sulfurated molecules (CS, OCS, SO, HCS$^{+}$ and
H$_{2}$S) relevant for the GEMS project; for these molecules, the chemical
variations are stronger for the 0.1 Myr model, especially for SO and OCS.
The observational effects will be largely discussed in the following sections.

Up to this point, we have presented the results of the MHD simulation,
which provides cubes of density, magnetic field and gas velocity field,
and the chemical post-processing with \textsc{Nautilus}, 
which provides cubes with the chemical abundances. However, we still need to perform
a last post-processing of these cubes to recover the observational imprint (spectral cubes), and 
with that purpose we used RADMC-3D \citep{2012ascl.soft02015D} v2.0.
RADMC-3D is a multi-purpose radiative transfer code for astrophysics that
allows, among other things, to create line radiative transfer
images and spectra of molecular lines, or in other words, synthetic
spectral cubes of rotational line emission. 
For each of the molecules studied in 
this work, we generate synthetic spectral cubes for only one
rotational transition that was observed during the GEMS programme; the full
list of rotational transitions and corresponding references for the collisional
coefficients is shown in Table \ref{tab:molecular_lines}. We use as input parameters
the chemical abundances of the desired species computed by \textsc{Nautilus}
assuming an isotopic ratio $^{12}$CO/$^{13}$CO $= 60$
\citep{2002ApJ...578..211S,2016ApJS..225...25G}
 for the isotopologues
$^{13}$CO, H$^{13}$CO$^{+}$ and H$^{13}$CN, and also assuming that the 
abundance of ortho-H$_{2}$S (o-H$_{2}$S, which is the simulated species) is 3/4 of the 
total abundance of H$_{2}$S \citepalias{GEMSIV}. The velocity field is taken from the MHD cube,
and we adopt a constant temperature 
of $T = 12$ K for consistency with our previous calculations. We
perform the line radiative transfer with the
Large Velocity Gradient + escape probability method available in RADMC-3D
\citep{2011MNRAS.412.1686S}, and retrieve as a general rule
three different views of the domain, two perpendicular to the magnetic field
and a third one where the magnetic field is parallel to the line of sight (one 
per cube face). Since
we have the full turbulent velocity field, we do not include any further widening
of the lines via the microturbulence parameter ($v_{\rm micro} = 0$). 
Finally, we assume an ortho-para H$_{2}$ ratio of 
$10^{-4}$ \citep{2022A&A...668A.131S} and choose a
velocity window with a typical width of 5 km s$^{-1}$ but for
a H$^{13}$CN, which presents hyperfine structure, 
for which we set a width of 25 km s$^{-1}$.

\begin{table*}
\footnotesize
      \caption{Rotational transitions considered for the line radiative transfer.}
      \label{tab:molecular_lines}
      \centering
      \begin{tabular}{c c c c c}
      \hline \hline
      Molecule  & Transition & Freq. & Beam size$^{\rm (a)}$ & Ref. collision coefs.$^{\rm (b)}$\\
				& & (GHz)   & (arcsec)    &                         \\
      \hline
      $^{13}$CO & $1 \rightarrow 0$ &  110.201 & 22.32 & \citet{2010ApJ...718.1062Y}\\
      H$^{13}$CO$^{+}$ & $1 \rightarrow 0$ & 86.754 & 28.36 & \citet{2014MNRAS.441..664Y} \\
      H$^{13}$CN & $1 \rightarrow 0$ & 86.340 & 28.49 & \citet{Navarro-Almaida2023} \\
      HNC & $1 \rightarrow 0$ & 90.664 & 27.13 & \citet{2011PCCP...13.8204D} \\
      CS & $2 \rightarrow 1$ & 97.981 & 25.11 & \citet{2006AA...451.1125L} \\
      OCS & $7 \rightarrow 6$ & 85.139 & 28.89 & \citet{1978ApJS...37..169G} \\
      SO & $(2,3) \rightarrow (1,2)$ & 99.300 & 24.77 & \citet{2007JChPh.126p4312L} \\
      HCS$^{+}$  & $2 \rightarrow 1$& 85.348 & 28.82 & \citet{1999MNRAS.305..651F} \\
      H$_{2}$S & $(\rm 1_{10} \rightarrow 1_{01})$ & 168.763 & 14.58 & \citet{2020MNRAS.494.5239D} \\
      \hline
      \end{tabular}
      \tablefoot{(a) Beam sizes are Half Power Beam Width (HPBW)
      computed from the IRAM 30 m efficiencies
      as HPBW = 2460/Freq. [GHz]. (b) Collision files have been retrieved
      from the CASSIS Collision Database (\url{http://cassis.irap.omp.eu/}), with the
      exception of the file for H$^{13}$CN. Rest frequencies in the table
      are taken from the collision files. Only collisions with 
      H$_{2}$ are relevant in a dark molecular cloud.}
\end{table*}

Apart from considering three different lines of sight for each simulation
in order to explore any effects of the projected magnetic field, it is also 
important to explore the influence of the hypothesis made on the molecular
abundances for the analysed transitions. When trying to fit an observed spectrum,
one usually has to assume an average gas density in the line of sight that
is translated into a chemical abundance via a radiative transfer programme 
such as RADEX \citep{2007A&A...468..627V}. Therefore, apart from the 
three-dimensional abundances obtained as explained at the beginning of this
section (which will be referred in the rest of the article as the 
`turbulent' abundances), we considered the case when the
gas density is taken as the average gas density ($n_{\rm H} = 5.14 \times 10^{4}$ cm$^{-3}$)
and the corresponding molecular abundance is computed with \textsc{Nautilus}, 
in other words, we are considering that the medium is uniform, an hypothesis
usually adopted when trying to fit observations. 



\section{Kinematics: Line profiles} \label{sec:line_profiles}

\begin{figure*}
\centering
\includegraphics[width = .33\textwidth]{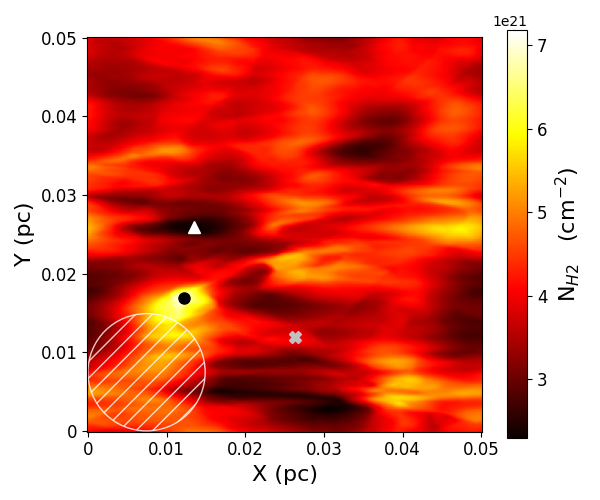}\includegraphics[width = .33\textwidth]{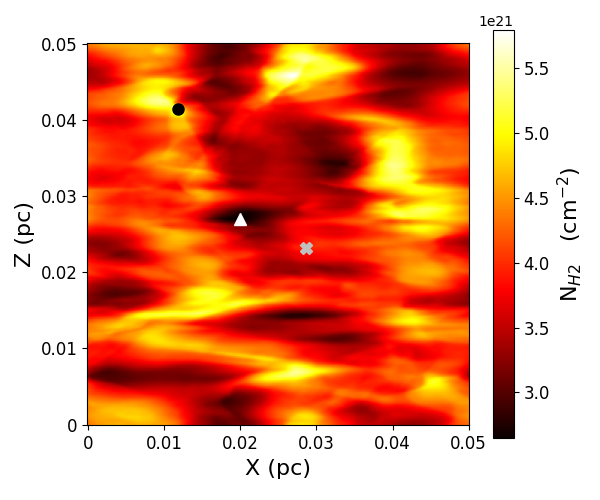}\includegraphics[width = .33\textwidth]{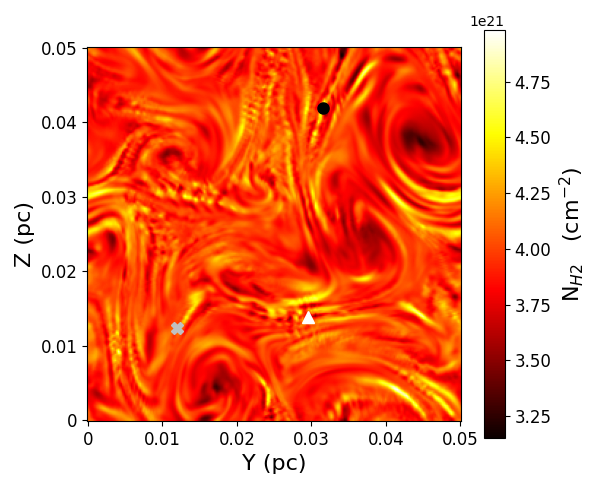}
\caption{Molecular hydrogen column density maps for the three faces of the simulation before convolution. 
Depending on the plane
chosen the value range varies slightly, being the yz plane the one with the less appreciable differences. 
In the first panel, a hatched white circle represents the IRAM 30 m beam
efficiency for $^{13}$CO ($1 \rightarrow 0$), which is the molecule used for the discussion
of the kinematics.
On each plot, the black
circle represents the line of sight with the higher column density, the white triangle the one with lower column density, and the
grey cross indicates the position where the column density is equal to the
mean value ($3.97 \times 10^{21}$ cm$^{-2}$), which corresponds
to a line of sight with a mean particle density of $n_{\rm H_{2}} = 2.17 \times 10^{4}$ cm$^{-3}$. The maximum
value of column density corresponds to an extinction
value of $A_{V} \sim 4$ mag, and the mean value is not unique: the fraction of the surface within the
10\% of this value is $\sim 45 \%$ in the xy plane, $\sim 49 \%$ in the xz plane, and $\sim 96 \%$ in the yz plane. 
These lines of sight have been chosen excluding the boundaries,
so that when convolving with the beam
highlighted in the first panel there is not any loss of information.}
\label{fig:column_density_maps}
\end{figure*}

First, we explored the influence of the relative alignment between
the line of sight (LoS) and the global magnetic field direction, which
is mostly parallel to the x-axis, on the predicted line profiles. 
For better illustrating this situation, in Fig. \ref{fig:column_density_maps}
we show the column density maps $N_{H_{2}}$ for the simulation on 
each of the three faces of the cube; two of the faces (xy, xz) are perpendicular to
the magnetic field direction, while the third face (yz) shows the column density
map when integrating along the direction of the mean magnetic field.
On each map, there are three individual lines of
sight represented by symbols: the black circle corresponds to the
line of sight with the highest column density, 
the white triangle to the lowest one, and the grey cross is the 
point of the map where the column density equals the mean column density. 
Also included in these plots is the IRAM beam size for $^{13}$CO ($1 \rightarrow 0$)
assuming
a nominal distance of 140 pc for the cloud, which is
the transition chosen for the discussion of the effects of turbulence on
the kinematics. We chose $^{13}$CO instead of $^{12}$CO for this discussion because
our observations as well as the post-processed simulations showed evidence
to be optically thick. Besides, we decided to focus this discussion on
$^{13}$CO instead of on other molecules because its emission was stronger, although
other transitions should present a similar behaviour.

In Fig. \ref{fig:LoS_MHD_IRAM} we show the line profiles for
$^{13}$CO ($1 \rightarrow 0$) -- assuming a chemical age of 1 Myr --
for these three lines of sight and for two faces of the cube,
one parallel and one perpendicular to the magnetic field, as predicted
by RADMC-3D for the simulation (label `MHD')
and after convolution with the IRAM 30 m beam (label `IRAM'); we
include in the discussion only one of the lines of sight perpendicular to the
magnetic field because the results in both cases are similar. From these
plots, it is evident that the relative alignment between the 
line of sight and the magnetic field makes a difference in the observed
line profiles, and in order to quantify these effects we computed
the second-order moment ($M_{2}$) of the integrated
intensity maps, which gives an estimate of the
spectral line width. This second-order moment is computed as:

\begin{figure*}
\centering
\includegraphics[width = .97\textwidth]{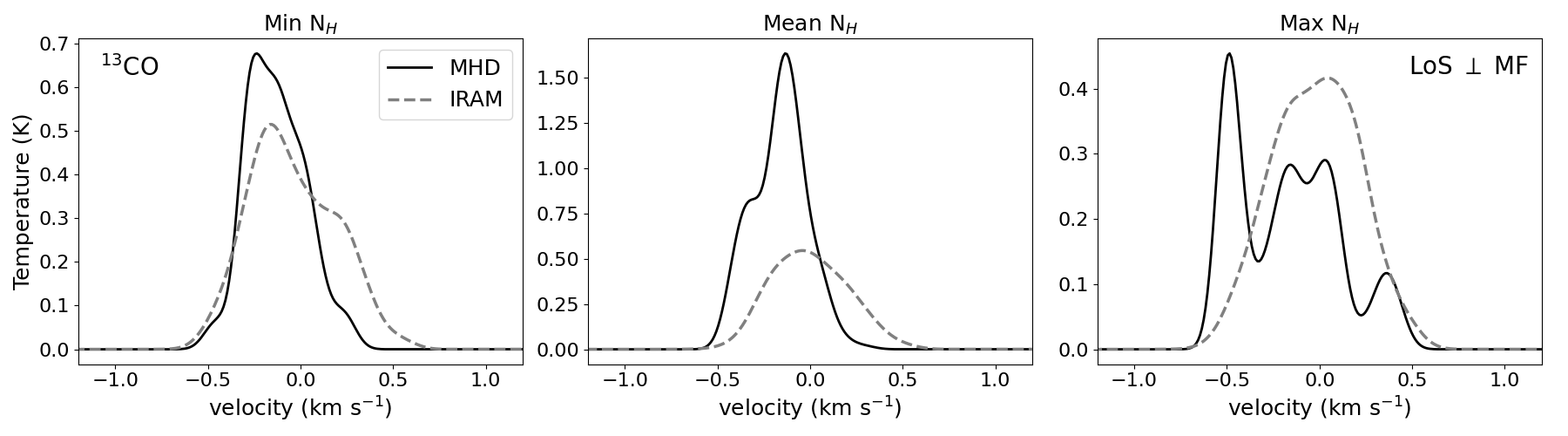}
\includegraphics[width = .97\textwidth]{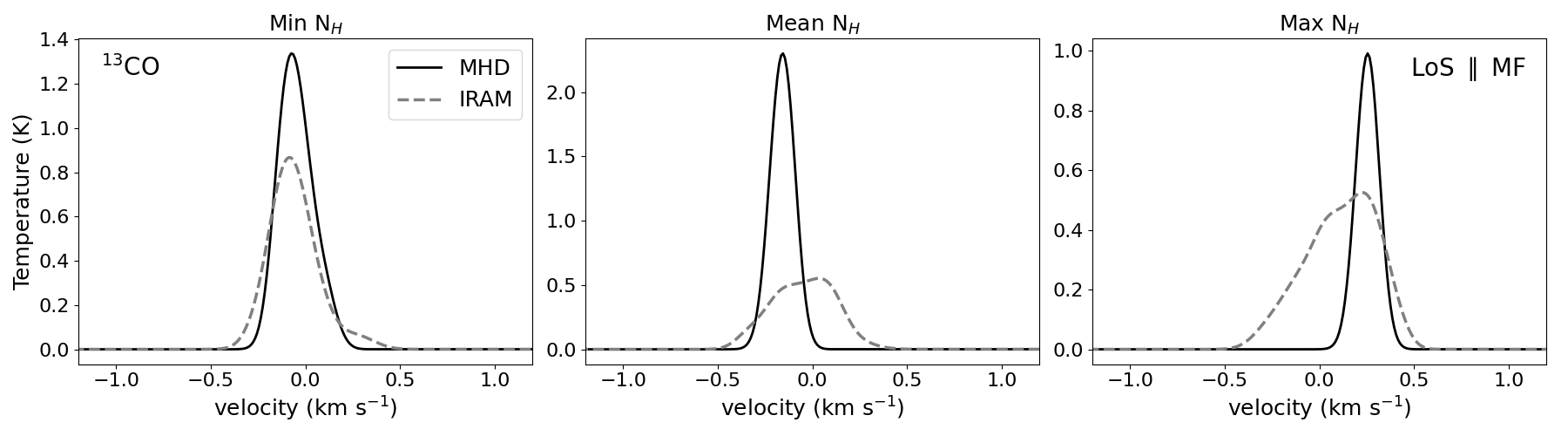}
\caption{Line profiles predicted by RADMC-3D for $^{13}$CO \trans{1}{0}
for two faces of the
cube and three lines of sight. The solid black line corresponds to the
post-processed simulation, which assuming a nominal distance of
$d = 140$ pc corresponds to a scale of 0.28\arcsec/cell; the dashed grey
line corresponds to the line profile convolved to the IRAM 30 m resolution
for the selected transition, which is 22.32\arcsec. The difference in 
velocity range for both the upper and lower panels, which corresponds to
lines of sight perpendicular (xy plane, left panel in Fig. \ref{fig:column_density_maps})
and parallel to the magnetic field (yz plane, right panel in  Fig. \ref{fig:column_density_maps})
respectively.}
\label{fig:LoS_MHD_IRAM}
\end{figure*}

\begin{equation}
M_{2} = \frac{\int T_{b}(v)(v - M_{1})^{2} dv}{\int T_{b}(v) dv}
\label{eq:second-order_moment}
\end{equation}

where $T_{b}$ is the temperature of the line in K, 
$v$ is the velocity in km s$^{-1}$, and
 $M_{1}$ is the first-order moment also in km s$^{-1}$:
 
\begin{equation}
M_{1} = \frac{\int v T_{b}(v)dv}{\int T_{b}(v) dv}
\label{eq:first-order_moment}
\end{equation}

By definition, the units of $M_{2}$ are km$^{2}$ s$^{-2}$, but in order
to provide a quantity in spectral units (km s$^{-1}$), 
we assume that $M_{2}$ represents the variance of a Gaussian distribution, and 
therefore
the dispersion can be computed as $\sigma = \sqrt{M_{2}}$, and the corresponding
line width (or full width at half maximum, FWHM) is  $2\sqrt{2 \ln 2}\sigma$.

\begin{table*}
\caption{Statistical measurements of the line widths for 
$^{13}$ CO \trans{1}{0} based on the
second-order moment map.}
\label{tab:linewidths}
\centering
\begin{tabular}{ c c c c|c c c c}
\hline \hline
     \multicolumn{4}{c}{LoS $\perp$ MF} & \multicolumn{4}{c}{LoS $\parallel$ MF} \\
     \multicolumn{2}{c}{MHD} & \multicolumn{2}{c}{IRAM} & \multicolumn{2}{c}{MHD} & \multicolumn{2}{c}{IRAM} \\
Median & IQR & Median & IQR & Median & IQR & Median & IQR \\
     & km s$^{-1}$ & km s$^{-1}$ & km s$^{-1}$ & km s$^{-1}$ & km s$^{-1}$ & km s$^{-1}$ & km s$^{-1}$\\
\hline
 0.4137 & 0.1227 & 0.4990 & 0.0732 & 0.1524 & 0.0174 & 0.3674 & 0.0815 \\
\hline
\end{tabular}
\tablefoot{Relative
alignments between the observer and the mean magnetic field (MF) direction considered
are parallel ($\parallel$) and perpendicular ($\perp$). The interquartile
range (IQR) provides the width of the interval that contains the central 50\% of
the data.
}
\end{table*}

In Table \ref{tab:linewidths}, we provide some statistical measurements of the
line widths of $^{13}$CO \trans{1}{0} for the two faces of the cube
shown in Fig. \ref{fig:LoS_MHD_IRAM}. As first appreciated from this
plot, the quantities in Table \ref{tab:linewidths} show that
in the `MHD' case the median line widths for lines of sight perpendicular
to the magnetic field are two to three times larger than 
for the case when the line of sight is
parallel to the magnetic field. However, this dependency of line width with
respect to the direction of the magnetic field might be lost when
observing with a telescope due to its lower resolution. For this discussion,
we assumed that our simulations correspond to a cloud located at
a distance of 140 pc, providing a spatial resolution of 0.28\arcsec; on
the other hand, the IRAM 30 m beam for the 
transition $^{13}$CO \trans{1}{0} is 22.32\arcsec, and we used this
value to degrade the post-processed maps to the IRAM 30 m resolution.
If we compare the median\footnote{This median is computed
taking into account all the image pixels, including the borders. However,
the differences in the median if borders are removed are of the order of
1\%.} line widths for the MHD and IRAM cases, we see that for the line of sight
perpendicular to the magnetic field, the IRAM line width is 1.21 times wider
than in the MHD case, while for line of 
sight parallel to the magnetic field
the difference between MHD and IRAM line widths 
is 2.41.
This is natural, since lines of sight parallel to the magnetic field
result in narrow, single line profiles for the MHD case
that suffer more from convolution,
while perpendicular lines of sight show a large number of discrete
components, and hence a wider line width; as a consequence,
convolution of lines of sight perpendicular to the magnetic field
results in a lower loss of information.
This difference in line structure depending on the simulation parameters
is also reflected in the position velocity diagrams, provided in
 Appendix \ref{appendix:PV_diagrams}. \par
 
In view of these results, it
is tempting to try to establish some criterion to determine
relative alignment with respect to the magnetic field based on the
observed line profiles. However, we want to note that although
promising, we cannot draw firm conclusions 
since we made many assumptions that 
limit our conclusions; a further discussion
on the limitations of this work is provider
later in the paper.
Nevertheless, we want to remark that
this line of research seems promising and we will properly address it
 in the future.


\section{Chemical abundances} \label{sec:abundances}

The main goal of this article is to determine to what extent
turbulence alone can produce significant variations on the 
chemistry that might explain the difficulties when fitting
observations of different molecules to a single 0D chemical
model. For that purpose, we post-processed our simulations
as explained in Sec. \ref{sec:chemistry} under the assumption
that the medium is uniform (label `UNIF') and post-processing
the simulation with the chemical networks for 0.1 Myr, similar to the time needed
to reach the stationary turbulent state, and
1.0 Myr (label `TURB'). In Sec.
\ref{sec:limitations} we will discuss the 
limitations and possible uncertainties
of this approach.
In practice, observations are
fitted assuming a single hydrogen density that may allow
to retrieve a single chemical abundance, so our `UNIF' case
corresponds to the procedure applied in our previous observational
works  (\citetalias{GEMSI, GEMSII, GEMSIV};
\citealt{Bulut2021, Spezzano2022, Esplugues2022}; \citetalias{GEMSVII}).\par

%

We worked with the integrated intensity
maps built from the RADMC-3D spectral cubes for each
case, which will be denoted as 
$I_{\rm UNIF}$ and
$I_{\rm TURB}$; the details of how these maps are obtained for each transition can be found
in Appendix \ref{appendix:RADMC3D_bkg}. 
In Fig. \ref{fig:relative_intensities_chemistry} we show
the box plot for the ratio of turbulent to
uniform integrated intensity maps for the case of the magnetic field perpendicular to the
line of sight. We show each molecule 
and transition listed in Table \ref{tab:molecular_lines} and
for the two chemical ages considered, 0.1 Myr (white boxes) and 1.0 Myr
(filled boxes). In this
plot, each box extends from quartile $Q_{1}$ (25\%) to
quartile $Q_{3}$ (75 \%), and the horizontal line inside the box
is the median value; the whiskers extend up to 1.5 the
interquartile range (IQR, $Q_{3} - Q_{1}$), and flier points have been
excluded from the plot in order to help the visualisation.

\begin{figure}
\centering
\includegraphics[width = \columnwidth]{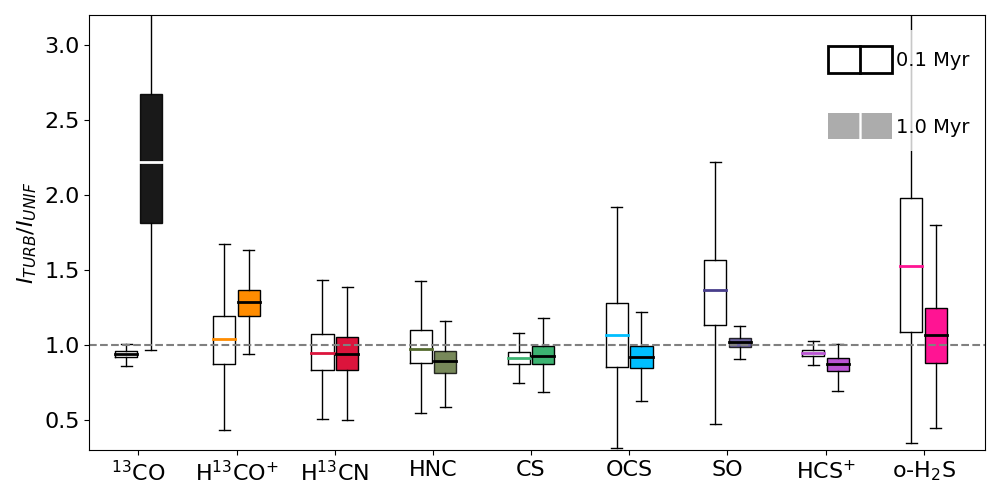}
\caption{Box plot for the ratio of integrated intensity maps 
$I_{\rm TURB} / I_{\rm UNIF}$ for the case of the magnetic field perpendicular to the
line of sight, where $I_{\rm TURB}$ is the
integrated intensity map for the full post-processed simulation
and $I_{\rm UNIF}$ is the intensity map assuming a constant
hydrogen density (and therefore constant
molecular abundance) along the line of sight.
For each
molecule we have two box plots: the left one corresponds to
the chemical network at
0.1 Myr, and the right one at 1 Myr. 
The y-axis has been scaled so that the main box
ranging from quartiles $Q_{1}$ (25 \%) to $Q_{3}$ (75 \%)
 is always shown and differences
among all molecules can be appreciated, although
in some cases the whiskers extend outside the plot.
This figure
corresponds to the case where the line of sight
is perpendicular to the mean magnetic field direction. An
analogous figure for the line of sight parallel to the
magnetic field is provided in Appendix \ref{fig5_additional}}
\label{fig:relative_intensities_chemistry}
\end{figure}

There are many conclusions that can be
drawn by inspection of Fig. \ref{fig:relative_intensities_chemistry}.
First, as we anticipated in the view of the predicted \textsc{Nautilus}
abundances for the density range and local properties
considered in this work (Fig. \ref{fig:chemical_models}),
the predicted line intensities for $^{13}$CO \trans{1}{0} depend strongly
on the chemical age. While for models with 0.1 Myr the
integrated intensities computed from the simulations are
slightly lower than those predicted under the uniform assumption, 
turbulence can only induce a slight dispersion on the integrated
intensity map that is very close to the uniform value. However,
at 1.0 Myr the predicted integrated intensities under the uniform
assumption are a factor of $\sim 2$ lower than those predicted for the
turbulent case (note the extension of the black filled box). 
It is known that the CO abundance is very sensitive to the chemical age in regions where the
dust temperature is below 15~K
\citep{Bulut2021}. This means that as long as the density is known, the CO 
abundance is a good chemical clock. Our results show that turbulence has a strong impact on
the mean CO abundance and therefore, on the observed line intensities. To ignore the fluctuations
that turbulence produces on the CO abundance would lead to an overestimation of the 
chemical age in dark clouds. \\

A similar striking behaviour is shown at 0.1~Myr for two
SO \trans{2, 3}{1, 2} and o-H$_{2}$S $(\rm 1_{10} \to 1_{01})$.
 As shown in Fig. \ref{fig:chemical_models}, the abundance of these sulfur-bearing species are very sensitive to
density, increasing by more than one order of magnitude.
This translates into integrated line intensities a factor of approximately 1.5 larger in the turbulent than in
the uniform case. These two species have been widely observed, especially in
warm regions \citep{Crockett2014, 2013AA...556A.143E, Elakel2022} and in bipolar
outflows \citep{Holdship2019, Shutzer2022}. Also abundant in dark clouds, chemical models usually failed to reproduce their abundances  
in these dense and cold regions (see e.g. \citetalias{GEMSII}). Our results show that
turbulence has a large impact on their chemical abundances, specially at chemical ages of a few 0.1 Myr  
which best reproduce the observations of less evolved starless cores \citep{Spezzano2022, Esplugues2022, 2022A&A...658A.168H}.
We recall that our model is isothermal, thus neglecting any heating of the gas due to
turbulence dissipation. The differences observed in the  SO \trans{2, 3}{1, 2} and o-H$_{2}$S $(\rm 1_{10} \to 1_{01})$
line intensities are only due to the variations of the density and kinematical structure in a turbulent filament.

The rest of the molecular species do not present significant
variations when comparing the turbulent and uniform cases, but it is
worth noting some minor dependencies. On the one hand, the integrated intensities
for H$^{13}$CO$^{+}$ \trans{1}{0} are roughly similar to the uniform case for 0.1 Myr,
but when fitting with a 1.0 Myr model integrated intensities for the turbulent
case are larger by a factor of $\sim 1.2$. On the other hand,
transitions such as H$^{13}$CN \trans{1}{0}, 
HNC \trans{1}{0}, CS \trans{2}{1}, and HCS$^{+}$ \trans{2}{1}
do not present 
strong dependencies on the chemical age, and the trend
of the medians is similar, what may indicate that a simultaneous
fitting of these four molecules to the same chemical model should
produce compatible results, although if we assume again a uniform
hydrogen density and molecular abundance the integrated
line intensities will be overestimated by a factor of $\sim 10-20$\%.\par

Up to this point, the discussion on the effects of turbulence on the
chemistry has been mainly qualitative based on the results displayed
in Fig. \ref{fig:relative_intensities_chemistry}. However, in Table
\ref{tab:relative_intensities_perpendicular} 
we provide medians and interquartile ranges for all the analysed transitions and
the two chemical ages considered, as well as for the post-processed
simulation (MHD) and the convolved maps (IRAM), for the
line of sight perpendicular 
and parallel to the magnetic
field direction. As already discussed from the plot, 
most of the integrated intensities are
a factor of $\sim 10$ \% larger
 when assuming a uniform gas 
distribution along the line of sight, with median values of the $I_{\rm TURB} / I_{\rm UNIF}$
ratio around 0.9. However, the main exceptions are
SO \trans{2,3}{1,2}, and o-H$_{2}$S $(\rm 1_{10} \to 1_{01})$ for the 0.1 Myr model,
and $^{13}$CO \trans{1}{0} for the 1.0 Myr model,
for which the post-processed
simulations indicate that the line intensities should be between a factor of 1.5
and 2 higher than when assuming a uniform gas density, with large dispersions.
One could argue that these statistical variations arising
from turbulence could be diluted when observing with a telescope, but as can be
seen from the comparison between the MHD and IRAM values in Table
\ref{tab:relative_intensities_perpendicular} 
the medians are equivalent; the main effect when degrading the resolution
to the beam size is the decrease of the interquartile range (the dispersion)
by an order of magnitude, which is quite significant but does not affect the main
results: turbulent motions in the gas may introduce a significant dispersion
in the chemistry that depends on chemical age and on the molecular species chosen. \par

\begin{table*}
\caption{Statistics of relative intensity maps 
$I_{\rm TURB} / I_{\rm UNIF}$
for the lines of sight considered in the
discussion.}
\label{tab:relative_intensities_perpendicular}
\centering
\begin{tabular}{c c|c c c c|c c c c}
\hline \hline
& & \multicolumn{4}{c}{0.1 Myr} & \multicolumn{4}{c}{1 Myr} \\
& & \multicolumn{2}{c}{LoS $\parallel$ MF} 
& \multicolumn{2}{c}{LoS $\perp$ MF} & 
\multicolumn{2}{c}{LoS $\parallel$ MF} 
& \multicolumn{2}{c}{LoS $\perp$ MF} \\ 
\hline
\multicolumn{2}{c}{Molecule} & Median & IQR & Median & IQR & Median & IQR & Median & IQR \\
&  & km s$^{-1}$ & km s$^{-1}$ & km s$^{-1}$ & km s$^{-1}$ & km s$^{-1}$ & km s$^{-1}$ & km s$^{-1}$ & km s$^{-1}$\\
\hline
$^{13}$CO  & MHD & 0.9729 & 0.0286 & 0.9389 & 0.0396 & 2.0681 & 0.9539 & 2.2228 & 0.8553 \\
\trans{1}{0}  & IRAM & 0.9654 & 0.0075 & 0.9350 & 0.0169 & 2.2478 & 0.1691 & 2.2653 & 0.2484 \\
  & & & & & & & & &\\
H$^{13}$CO$^{+}$   & MHD & 1.0859 & 0.1256 & 1.0388 & 0.3208 & 1.2501 & 0.1137 & 1.2884 & 0.1755 \\
\trans{1}{0}  & IRAM & 1.0941 & 0.0147 & 1.0398 & 0.0485 & 1.2489 & 0.0109 & 1.2850 & 0.0261 \\
  & & & & & & & & &\\
H$^{13}$CN  & MHD & 0.9890 & 0.1667 & 0.9471 & 0.2404 & 0.9964 & 0.3839 & 0.9413 & 0.2207 \\
\trans{1}{0}  & IRAM & 0.9978 & 0.0187 & 0.9680 & 0.0272 & 0.9844 & 0.0517 & 0.9442 & 0.0332 \\
  & & & & & & & & &\\
HNC  & MHD & 1.0065 & 0.3464 & 0.9753 & 0.2204 & 0.9413 & 0.2168 & 0.8921 & 0.1499 \\
\trans{1}{0}  & IRAM & 1.0296 & 0.0546 & 0.9917 & 0.0269 & 0.8959 & 0.0379 & 0.8781 & 0.0301 \\
  & & & & & & & & &\\
CS  & MHD & 0.9419 & 0.1074 & 0.9149 & 0.0824 & 0.9627 & 0.2221 & 0.9292 & 0.1232 \\
\trans{2}{1}  & IRAM & 0.9221 & 0.0215 & 0.9147 & 0.0233 & 0.9576 & 0.0405 & 0.9333 & 0.0278 \\
  & & & & & & & & &\\
OCS  & MHD & 1.0746 & 0.1502 & 1.0638 & 0.4266 & 0.9281 & 0.0747 & 0.9182 & 0.1480 \\
\trans{7}{6}  & IRAM & 1.0737 & 0.0216 & 1.0744 & 0.0647 & 0.9184 & 0.0099 & 0.9206 & 0.0243 \\
  & & & & & & & & &\\
SO  & MHD & 1.3869 & 0.3195 & 1.3703 & 0.4358 & 1.0235 & 0.0671 & 1.0212 & 0.0544 \\
\trans{2,3}{1,2}  & IRAM & 1.4039 & 0.0446 & 1.3493 & 0.0828 & 1.0028 & 0.0100 & 1.0137 & 0.0145 \\
  & & & & & & & & &\\
HCS$^{+}$  & MHD & 0.9548 & 0.0640 & 0.9497 & 0.0387 & 0.8798 & 0.1086 & 0.8720 & 0.0896 \\
\trans{2}{1} & IRAM & 0.9425 & 0.0067 & 0.9450 & 0.0079 & 0.8653 & 0.0133 & 0.8681 & 0.0176 \\
  & & & & & & & & &\\
o-H$_{2}$S  & MHD & 1.5179 & 0.4827 & 1.5284 & 0.8928 & 1.0989 & 0.1279 & 1.0681 & 0.3678 \\
$(\rm 1_{10} \to 1_{01})$ & IRAM & 1.5780 & 0.1696 & 1.5809 & 0.3735 & 1.1102 & 0.0422 & 1.0769 & 0.1437 \\
\hline 
\end{tabular} 
\tablefoot{For each molecule, we show medians and interquartile ranges
computed from the original maps
(label `MHD') and from the convolved maps (label `IRAM').}
\end{table*}

\section{Comparing with single-dish observations} \label{sec:discussion}
%
So far, we have focussed the discussion on the theoretical results.
However, it is also important to highlight the impact that our study
may have on the observations. With that purpose, we selected
a subsample of the positions observed within the GEMS project
in Taurus: 3 cuts in TMC-1 and 9 in B213
\citepalias{GEMSIV}.
The cuts were selected based on their estimated temperature (around
12 K) and molecular hydrogen densities (a few 10$^{4}$ cm$^{-3}$). As an
additional constraint, we imposed a non-null detection of 
H$^{13}$CN \trans{1}{0} in order to be able to do a more 
complete and meaningful comparison. After applying these criteria, we
ended up with a sample of 37 positions with non-null detections
for essentially all the transitions. The main exception is
OCS \trans{7}{6}, a transition for which we do not detect emission for 19 positions;
additionally, two positions are lacking o-H$_{2}$S $(\rm 1_{10} \to 1_{01})$ emission, 
and another four do not present emission of the HCS$^{+}$ \trans{2}{1} line. \par

\begin{figure*}[h!]
\centering
\includegraphics[width=.45\textwidth]{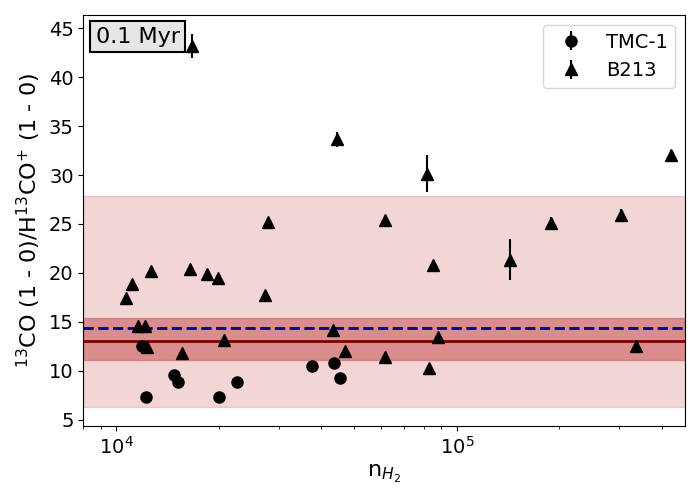}\includegraphics[width=.45\textwidth]{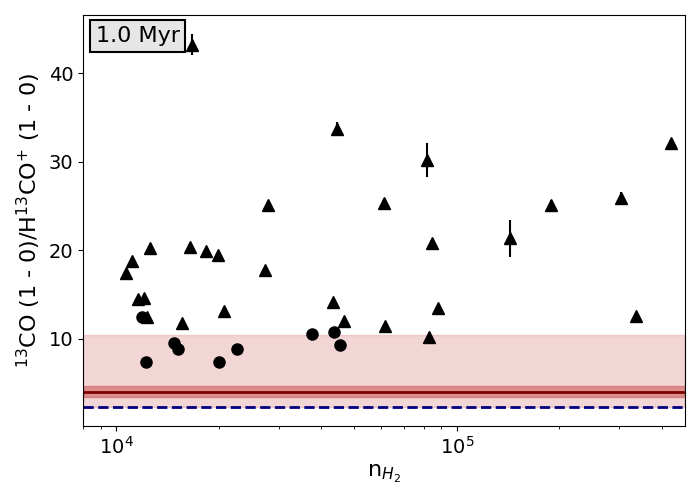}
\includegraphics[width=.45\textwidth]{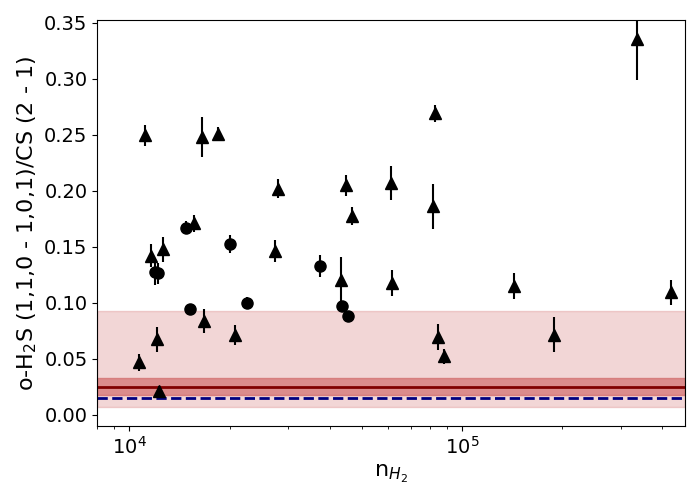}\includegraphics[width=.45\textwidth]{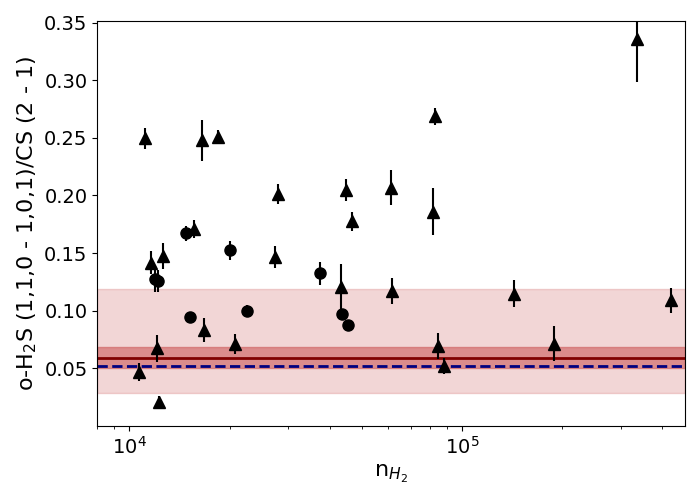}
\includegraphics[width=.45\textwidth]{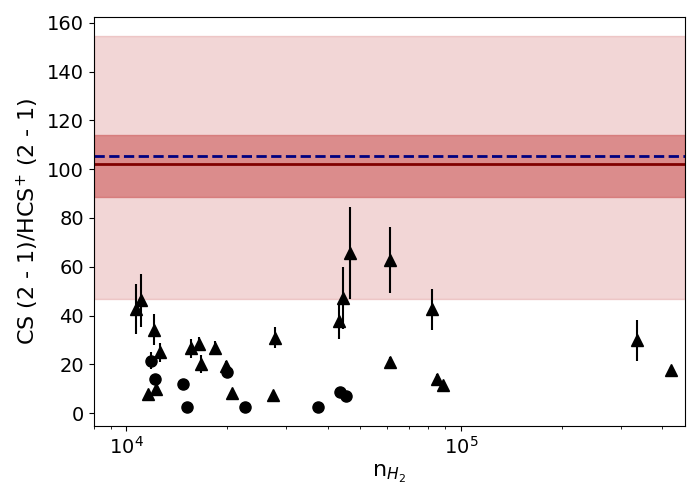}\includegraphics[width=.45\textwidth]{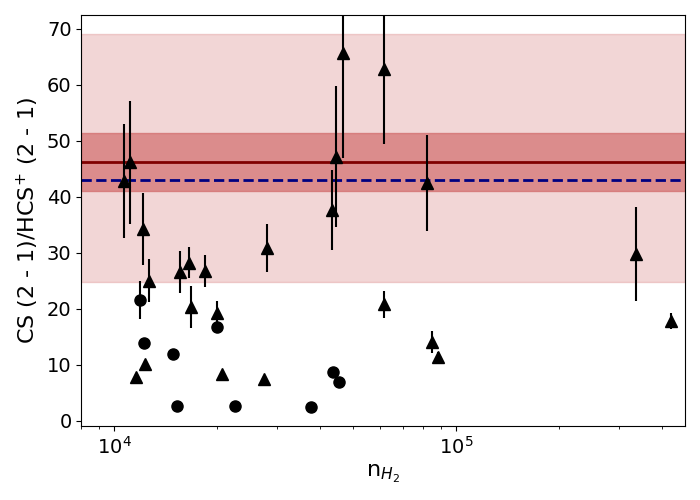}
\caption{Comparison between predicted and observed values of the ratios 
$^{13}$CO \trans{1}{0} / H$^{13}$CO$^{+}$ \trans{1}{0}, 
o-H$_{2}$S $(\rm 1_{10} \to 1_{01})$ / CS \trans{2}{1}, and 
CS \trans{2}{1} / HCS$^{+}$ \trans{2}{1}. The left panels correspond to
the post-processed simulations at 0.1 Myr while the right panels correspond
to 1.0 Myr. On each panel, the median value for the turbulent (uniform)
case is highlighted by the solid (dashed) line, the IQR is represented by the 
dark area and the full range (from minimum to maximum) is represented by
the light-coloured area. The uniform and turbulent cases are
coded in blue and red, respectively. TMC-1 points are represented by circles and 
B213 by triangles, and error bars are always plotted although in most
cases they are smaller than the marker. The error bars correspond
to the statistical errors from fitting the lines and do not include
systematics \citepalias{GEMSIV}.}
\label{fig:comparison_gems_M0}
\end{figure*}

\begin{figure*}[h!]
\centering
\includegraphics[width=.45\textwidth]{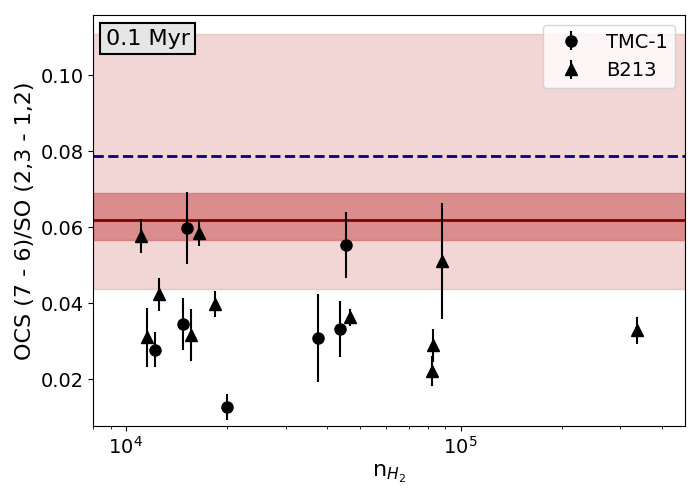}\includegraphics[width=.45\textwidth]{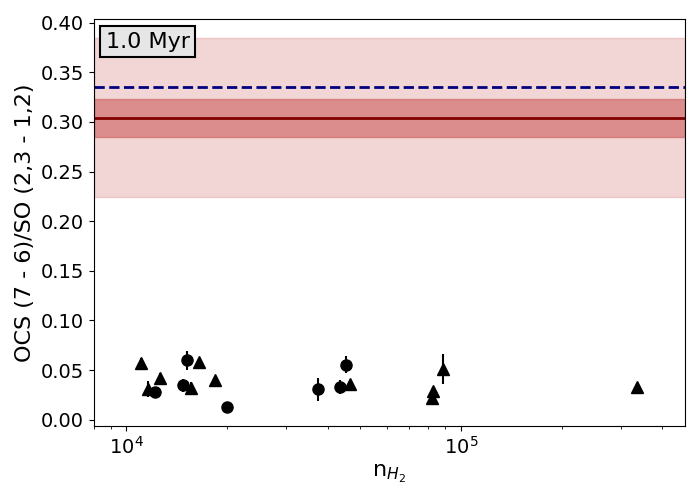}
\includegraphics[width=.45\textwidth]{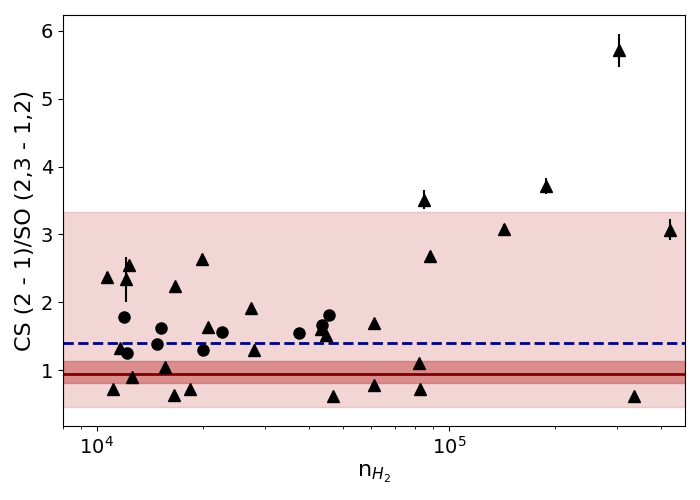}\includegraphics[width=.45\textwidth]{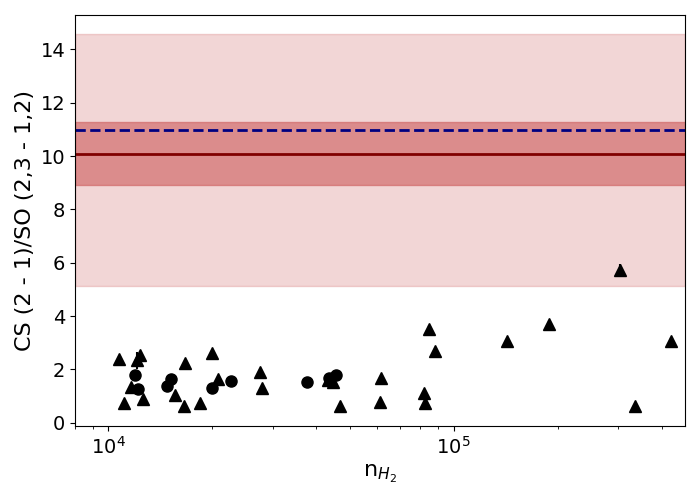}
\includegraphics[width=.45\textwidth]{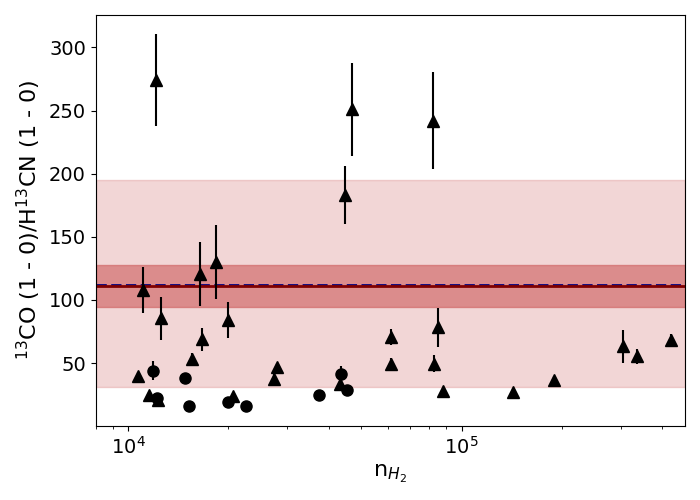}\includegraphics[width=.45\textwidth]{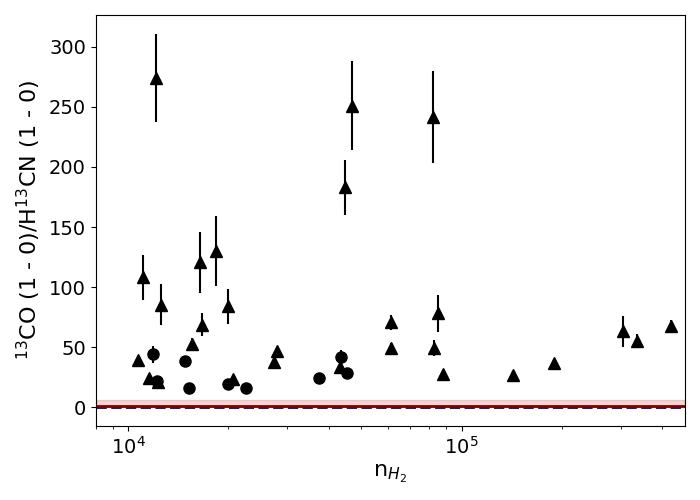}
\includegraphics[width=.45\textwidth]{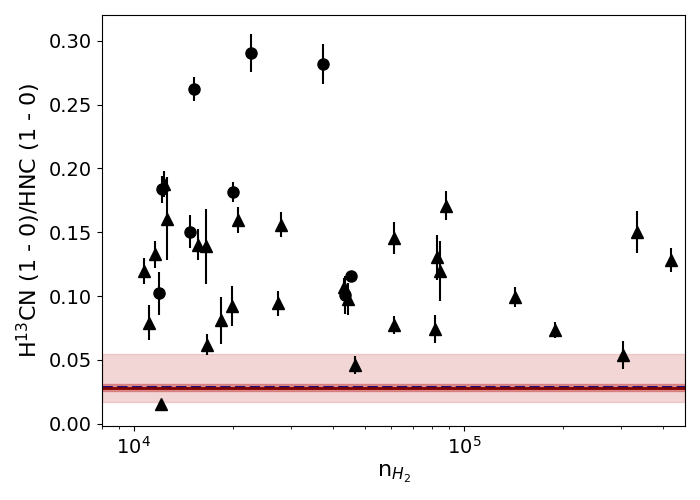}\includegraphics[width=.45\textwidth]{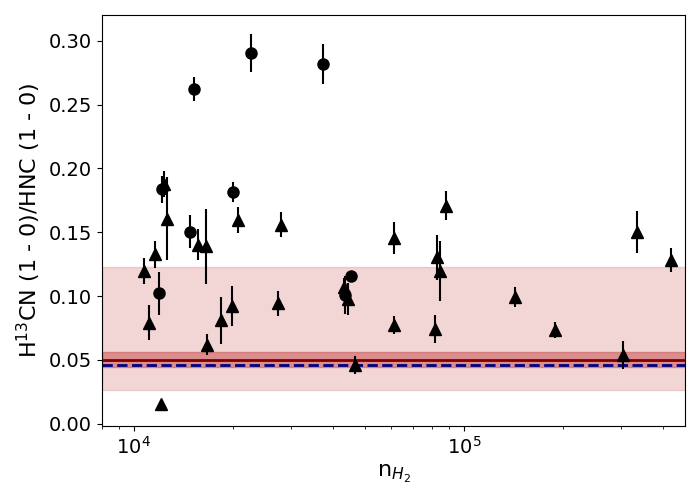}
\caption{Same as Fig. \ref{fig:comparison_gems_M0} but for the ratios
OCS \trans{7}{6} / SO \trans{2,3}{1,2},
CS \trans{2}{1} / SO \trans{2,3}{1,2},
$^{13}$CO \trans{1}{0} / H$^{13}$CN \trans{1}{0}, and
H$^{13}$CN \trans{1}{0} / HNC \trans{1}{0}.
}
\label{fig:comparison_gems_M0_2}
\end{figure*}

A direct comparison between our synthetic observations and real data is
not easy because our simulation box represents a fraction of the cloud while observations
account for emission from all the material along the line of sight. For that
reason, we opted for comparing integrated line intensity ratios that are less sensitive to
the length of the line of sight in the optically thin case
and that are commonly used by
observational astronomers
(\citetalias{GEMSI}, \citetalias{GEMSVII}, \citealt{2018A&A...617A..28S, 2019ApJ...880..127J,2020A&A...635A...4H,2023A&A...672A..96G}):
$^{13}$CO \trans{1}{0} / H$^{13}$CO$^{+}$ \trans{1}{0}, 
o-H$_{2}$S $(\rm 1_{10} \to 1_{01})$ / CS \trans{2}{1},
CS \trans{2}{1} / HCS$^{+}$ \trans{2}{1},
OCS \trans{7}{6} / SO \trans{2,3}{1,2},
CS \trans{2}{1} / SO \trans{2,3}{1,2},
$^{13}$CO \trans{1}{0} / H$^{13}$CN \trans{1}{0}, and
H$^{13}$CN \trans{1}{0} / HNC \trans{1}{0}. 
In the case of optically thin emission, the integrated line intensity is  proportional to the column density 
and excitation temperature of a given species in
a given interval of the projected velocity along the line of sight. Therefore, it depends on the density and 
kinematical structure of the cloud as well as on the physical conditions of the emitting
gas (i.e. its temperature). Trying to evaluate the impact of the turbulence on the chemistry, we considered: (i) the uniform case with constant density and molecular abundances explained above; and
(ii) turbulent case with the densities and the chemical abundances obtained in Sect.~\ref{sec:abundances}.
In the two cases, we used the velocity field provided  when considered that the line of sight is perpendicular
to the magnetic field. We are aware that some of the selected lines might be optically thick
towards some positions. In order to minimise this problem
we used the $^{13}$C isotopologues for HCN and HCO$^+$.  
The comparison between 
the output of these simulations and GEMS observations
is made under a descriptive approach since our observational sample is quite limited in size.
For the sake of interpretation, we computed the credible regions for each simulation
with a credibility level of 95\% and counted the number of observational
points that lay inside them, without taking into account the observational errors
that arise from the line fitting process; the results are listed in Table \ref{tab:credible_regions}. 
These regions are, in practice, intervals which contain
the data corresponding to our synthetic populations with a probability of 95\%, and are the
bayesian equivalent to confidence intervals. As a complementary visual support, we computed
the median and interquartile range of these synthetic distributions and the results, together
with the observational measurements, are shown
in Figures \ref{fig:comparison_gems_M0} and \ref{fig:comparison_gems_M0_2},
where we plot the predicted values for the uniform and turbulent cases
coded in blue and red respectively together with the GEMS integrated
intensities towards TMC-1 (circles) and B213 (triangles).

\begin{table*}
\caption{Credible regions (95 \%) derived from our synthetic data for the turbulent
and uniform cases and percentage of observational points contained in the interval.}
\label{tab:credible_regions}
\centering
\begin{tabular}{c c|c c| c c}
\hline \hline
& & \multicolumn{2}{c}{0.1 Myr} 
& \multicolumn{2}{c}{1.0 Myr}  \\ 
\hline
\multicolumn{2}{c}{Ratio} & Credib. Region & Obs. points (\%) & Credib. Region & Obs. points (\%) \\
\hline
$^{13}$CO/H$^{13}$CO$^{+}$  & Unif & (10.61, 13.01) & 18.92 & (1.82, 1.86) & 0 \\
\trans{1}{0} \trans{1}{0}  & Turb & (7.29, 15.78) & 54.05 & (2.04, 4.53) & 0 \\
  & & & & &\\
o-H$_{2}$S/CS  & Unif & (0.01, 0.02) & 0 & (0.04, 0.07) & 5.71 \\
\trans{1,1,0}{1,0,1} \trans{2}{1}  & Turb & (0.01, 0.05) & 5.71 & (0.04, 0.09) & 20 \\
  & & & & &\\
CS / HCS$^{+}$  & Unif & (73.87, 132.91) & 0 & (30.92,53.07) & 18.18 \\
 \trans{2}{1} \trans{2}{1}  & Turb & (71.29, 131.44) & 0 & (32.85, 58.58) & 18.18 \\
  & & & & &\\
OCS / SO  & Unif & (0.07, 0.09) & 0 & (0.33, 0.34) & 0 \\
 \trans{7}{6} \trans{2,3}{1,2}  & Turb & (0.05, 0.08) & 27.78 & (0.25, 0.35) & 0 \\
  & & & & &\\
CS / SO  & Unif & (1.12, 1.7) & 35.14 & (8, 13.5) & 0 \\
 \trans{2}{1} \trans{2,3}{1,2}  & Turb & (0.55, 1.64) & 56.76 & (7.15, 13.03) & 0 \\
  & & & & &\\
$^{13}$CO / H$^{13}$CN  & Unif & (73.16, 153.2) & 13.51 & (0.11, 0.56) & 0 \\
 \trans{1}{0} \trans{1}{0}  & Turb & (60.23, 156.15) & 24.32 & (0.16, 0.68) & 0 \\
  & & & & &\\
H$^{13}$CN/HNC  & Unif & (0.029, 0.031) & 0 & (0.035, 0.07) & 13.51 \\
 \trans{1}{0} \trans{1}{0}  & Turb & (0.025, 0.041) & 0 & (0.042, 0.9) & 24.32 \\
\hline 
\end{tabular} 

\end{table*}

In agreement with our previous fits \citepalias{GEMSVII}, most of the line integrated intensity ratios are
better fitted with t=0.1~Myr. At this chemical time, the difference between the median of the 
$^{13}$CO/H$^{13}$CO$^+$, CS/HCS$^+$, $^{13}$CO/H$^{13}$CN, and H$^{13}$CN/HNC values obtained in the uniform and turbulent cases are shallow. 
This is expected taking into account the results shown in Fig.~\ref{fig:relative_intensities_chemistry}. The differences in the integrated line intensities of
these transitions between the uniform and turbulent cases are less than 50\%. However, we would like to comment that this difference goes in the direction of
improving the agreement between simulations and observations,
indicating that our approach goes in the right direction. 
Moreover, this improvement is more significative if we consider the range of values (shaded red area in Fig.~\ref{fig:comparison_gems_M0} and \ref{fig:comparison_gems_M0_2} ) obtained in our synthetic grid,
as well as the credible regions in Table \ref{tab:credible_regions}. For instance,
all the positions in TMC-1 and a large fraction of those in B 213 lie in the shaded red area of the $^{13}$CO/H$^{13}$CO$^+$ panel, which accounts for the
54.05 \% of the observational data.
There are three cases in which we found a large difference (larger than 50\%) between the uniform and turbulent simulations.
These are  o-H$_2$S/CS, OCS/SO, and CS/SO. In the case of  o-H$_2$S/CS, the inclusion of turbulence definitely improves
the agreement with Taurus observations, in the sense that 
more observational points are contained in the credible regions for the
turbulent case than for the uniform case, but in the view of the plots
we do not obtain a good result yet
because the median value is still quite far from the centre of the cloud of observational
points.  We do obtain a reasonable agreement
with observations in the case of the OCS/SO ratio when considering the turbulent case at early ages (0.1 Myr). 
In the case of the CS/SO ratio, the dispersion in
the observed values hinders to establish any conclusion,
but taking turbulence into consideration increases the number of points
inside the credible region from 35.14 \% for the uniform case
up to 56.76 \% for the turbulent one. \par

Only the CS/HCS$^+$ and o-H$_2$S/CS ratios are better 
explained using t = 1~Myr. This difficulty to explain the observations of all species
using the same chemical time has been previously discussed (see e.g. \citealp{Bulut2021}, 
\citetalias{GEMSVII} and \citealp{Taillard2023}). Indeed, in \citetalias{GEMSVII} we already proposed that
the positions located at low extinction are better fitted with t = 1~Myr while those at higher 
extinctions (and densities) were fitted with shorter chemical times. \citet{Taillard2023} obtained a similar result
based on methanol observations and suggested that this might be related with the dynamical
evolution of the cloud contraction which is accelerated as the density increase.
Regarding this interpretation, it is interesting to comment
that HCS$^+$ and H$_2$S can be considered tracers of the most diffuse part
of the gas since their emission is
dominated by the outer layers of the cloud \citepalias{GEMSI, GEMSII}.
We would also like to mention the behavior of the $^{13}$CO/H$^{13}$CO$^+$ ratio, which 
increases a factor of $\sim$2 in the turbulent case compared with the uniform one. Still,
the values predicted by our simulations are far from those observed in Taurus, 
and this is
a significant difference that should be taken into account.

As a final conclusion, it is evident that, when including
turbulence, model predictions approach better the observational values. The 
predicted median for the turbulent case always follows better the mean
observational value than the uniform case, and the range it spans is
always broader. This difference is specially important for the sulfur-bearing species
SO, OCS, and o-H$_2$S at early times, and for CO  at ages later than a few 0.1 Myr.
As a consequence, it is important to take into
account turbulence when fitting the observations of these species, for instance
via numerical modelling.


\section{Limitations of the model} \label{sec:limitations}

The scope of this paper was to assess if
small scale structure not resolved in 
single-dish molecular observations could be
of importance when fitting the data, and how
biased the results are if a uniform medium is
assumed. Therefore, we built a very simple model
that responded to our needs as a first approximation,
but we want to remark that there are several limitations
that we disregarded for the sake of the
discussion but that could be biasing our 
interpretation. \par

First of all, the numerical model presented here
has not enough spatial resolution to correctly
analyse the properties of the turbulence. Since
we are considering quite a strong magnetic
field (40 $\mu$G), \citet{2009ApJ...691.1092L} showed
that for the turbulence to fully converge, we should
reach a resolution of at least 512$^{3}$. However, 
analysing the small-scale properties of the 
turbulence is out of the scope of this paper,
and we refer the reader to the recent work by
\citet{2023ApJ...942L..34G} for a deeper discussion
on that topic. In our case, we confirmed that
increasing the resolution from $256^{3}$ to $512^{3}$
resulted in smoother density and velocity PDFs, but
the general behaviour of the fluid is the same:
the PDFs become stable after two crossing times
and the differences in density distribution
would be important if we were analysing
the chemical evolution of the cloud at the same
time as its dynamical evolution, but that is
not the case. Actually, these differences in density
would be lost in the step of degrading the 
maps at the IRAM resolution, since an IRAM beam
contains $\sim 85^{3}$ cells,
so we expect the comparison with
observations not to be significantly biased due to the
numerical resolution, although the
analysis of the original MHD simulation is
likely
biased towards large-scale density fluctuations.

Then, we are considering separately
the dynamical and chemical evolution of the
cloud. In fact, we evolve our numerical model
until a stage that would correspond to a physical
age of 0.35 Myr, but then we post-processed
this density cube with chemical models
at 0.1 Myr and 1 Myr. We acknowledge that
chemical stability is reached later than
dynamical stability in a
turbulent box (López-Dones et al. in prep)
but, on the other hand, since we are post-processing
a stable density PDF 
with chemical models and the density
in our simulations is not affected by the chemistry,
it is reasonable to consider both
chemical models.

Another limitation could arise from the
LVG approximation for computing the molecular
emission. With this approximation, we are only
taking into account the density and velocity
distribution of the gas, but more sophisticated
methods could be followed to consider other
effects appart from non-LTE (see for
instance the recent work by
\citealt{2006A&A...445..591H} and references
therein). However, given the moderate column densities
considered in this work and provided we did not
observe any saturation of the lines for the chosen transitions, 
we assume the LVG
approximation to hold for our analysis.

Finally, we want to mention that even if
we are working with a limited resolution at which
the turbulence is not fully resolved, we also
explored other cases with different
turbulent mixing modes (purely solenoidal) and magnetic field
strength (90 $\mu$G) at the same resolution as the model
analysed in the main core of the paper. We did
not find any strong differences in the density
PDFs, and the same trend of narrower
line widths for lines of sight parallel to the
magnetic field was observed in all cases, with
increasing line widths at lines of sight perpendicular
to the magnetic field as we considered stronger
magnetic fields.\par

All in all, even if we are
working with a low-resolution model we do not
expect that the comparison with observations
performed in Sec. \ref{sec:discussion} will be
inconsistent, since we are comparing with
single-dish observations. However, we aim to develop
more realistic models in the future where we
evolve the chemistry together with the MHD evolution at a resolution
as high as possible that could serve to
define new constraints when fitting observations,
especially for the analysis of future interferometric data.


\section{Conclusions} \label{sec:conclusions}

In this work, we present a study on the influence of turbulence
on the chemical evolution of a molecular cloud,
focussing our analysis in the density fluctuations and velocity
fields induced by such turbulence. Based on a numerical
MHD simulation that reproduces the steady-state
of a transonic, turbulent molecular filament before collapse,
we derive chemical abundances with the \textsc{Nautilus} chemistry code
and retrieve synthetic maps of rotational emission for some molecules
observed during the IRAM 30 m Large Programme GEMS. We explore a turbulent model with
both compressive and solenoidal modes and a magnetic
field strength of
40 $\mu$G, as well as two
chemical ages (0.1 Myr and 1.0 Myr), and we analyse the synthetic emission maps
based on the relative alignment between the observer and the mean
magnetic field direction. Our main results can be summarised
as follows:

\begin{itemize}

\item When the 3D cubes of densities, velocities and abundances
are post-processed with RADMC-3D for obtaining synthetic
molecular line emission maps, even after convolution
with the IRAM 30 m beam 
we observe that the line profiles for
lines of sight perpendicular to the magnetic
field are wider than in the case when it is
parallel.

\item  In our analysis, we computed the chemical abundances
using \textsc{Nautilus} under two assumptions: that the medium is
homogeneous in density along the line of sight, and that the density
distribution is turbulent as given by the simulations.
 When turbulence is taken into account, the predicted integrated
intensities are different from those derived assuming a uniform gas density
along the line of sight. The differences are critical for some sulfur-bearing
molecules (SO, H$_{2}$S) at early ages and $^{13}$CO at 1.0 Myr 
(Fig. \ref{fig:relative_intensities_chemistry}).

\item When comparing our predictions with real observations of Taurus, we find
that the inclusion of turbulence always improves the results, and 
most of the analysed molecular ratios are compatible with the 
model at 0.1 Myr, the observationally 	derived chemical age 
\citepalias{GEMSVII}.

\end{itemize}

We conclude that turbulence may induce density variations which
play an important role in shaping
the chemistry of molecular clouds, especially for those molecules
sensitive to hydrogen density -- mainly CO and some sulfur-bearing
species in the cases analysed here. Therefore, turbulence should be
taken into account when trying to derive chemical abundances from
observations, using for instance numerical simulations as the ones
performed here with parameters relevant to those of the particular cloud
under study. Other effects, such as introducing a pre-phase model
or considering mixed dust populations of silicate and graphite grains might be
relevant and will be explored in future works.


\begin{acknowledgements}
      We want to thank the referee for their comments that helped
      to improve the clarity of this paper.
      L.B.-A. acknowledges the receipt of a Margarita Salas postdoctoral
      fellowship from Universidad Complutense de Madrid (CT31/21), funded by
      'Ministerio de Universidades' with Next Generation EU funds. L.B.-A. 
      also wants to thank Chang-Goo Kim for useful discussions on the general
      use of Athena++.
     A.F. thanks the Spanish MICIN for funding support through grant PID2019-106235GB-I00 and the European Research 
      Council (ERC) for funding under the Advanced Grant project SUL4LIFE, grant agreement No 101096293.
      R.L.G. would like to thank the "Physique Chimie du Milieu Interstellaire" (PCMI)
       programs of CNRS/INSU for there financial supports.
      A.I.G.C. and L.B.-A. acknowledge financial support through grant
      MINECO – PID2020-116726RB-I00.
      This research has made use of the computational facilities of the Joint
Center for Ultraviolet Astronomy in the campus of the Universidad
Complutense de Madrid, managed by J.C. Vallejo and R. de la Fuente Marcos.
      D.N.-A. acknowledges funding support from Fundación Ramón Areces through
       their international postdoc grant program.
       J.E.P. is supported by the Max-Planck Society. RMD has received funding 
       from "la Caixa" Foundation, under agreement LCF/BQ/PI22/11910030
\end{acknowledgements}

%
\bibliographystyle{aa} 
\bibliography{references} 
%

\begin{appendix}

\section{Influence of the initial conditions on the chemistry}\label{appendix:prephase_chemistry}

In this work, we take all of the chemical species to be initially
in atomic phase, except for hydrogen that is all locked into H$_{2}$. These
conditions are usually considered valid when analysing the chemical abundances
in molecular clouds (see e.g. \citealt{2018MNRAS.474.5575V, 2021A&A...647A.172W, 2023A&A...673A..34A}), but we also considered
the possibility of starting from a more complex medium as in \citet{2022A&A...657A.100C}.
With that purpose, we run a pre-phase model starting from the same
abundances listed in Table \ref{tab:initial_abundances}, assuming a
hydrogen density of $n_{\rm H} = 2 \times 10^{3}$ and the rest of the initial
conditions as in the model detailed in Sec. \ref{sec:chemistry}.
We follow the evolution of this lower-density medium until 1 Myr,
a stage where we have many molecules formed, and take these
abundances as the initial conditions for our test case. As a 
reference, we list in Table \ref{tab:initial_abundances_prephase}
the 10 most abundant molecules and the main sulfur reservoirs
of this prephase model, and the
full list of 911 species is available as supplementary
material.

\begin{table}
\centering
\caption{Initial elemental abundances with respect to H for the 
10 most abundant species of the 
pre-phase model plus the main sulfur reservoirs.}
\label{tab:initial_abundances_prephase}
\begin{tabular}{cc}
\hline \hline
Species & Abundance \\
\hline
H$_{2}$ & $4.98 \times 10^{-1}$ \\
He & 0.09 \\
H & $2.85 \times 10^{-3}$\\
CO & $9.6913\times 10^{-5}$\\
O & $6.7817\times 10^{-5}$ \\
K-H$_{2}$O & $4.9704\times 10^{-5}$\\
N & $3.8677 \times 10^{-5}$\\
C &  $2.6894 \times 10^{-5}$\\
K-HCN & $8.7744 \times 10^{-6}$ \\
K-CO2 & $7.482 \times 10^{-6}$ \\
S$^{+}$ & $5.067 \times 10^{-7}$ \\
S & $1.8112 \times 10^{-7}$ \\
\hline
\end{tabular}
\tablefoot{The prefix K- refers to molecules locked in the bulk of the dust grains, below the ice-coated surface.}
\end{table}

\begin{figure}[h!]
\centering
\includegraphics[width = \columnwidth]{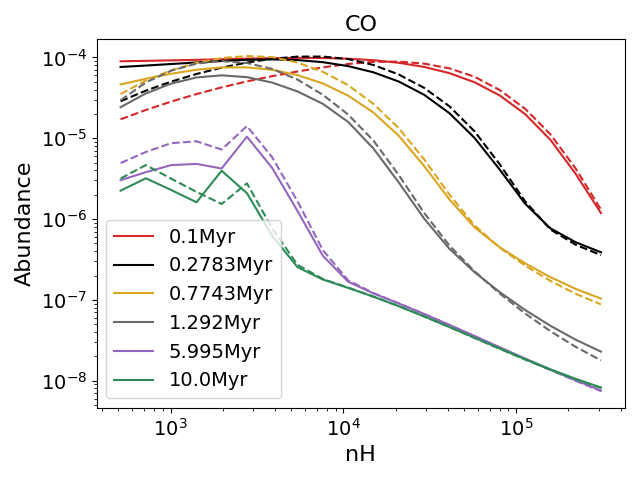}
\includegraphics[width = \columnwidth]{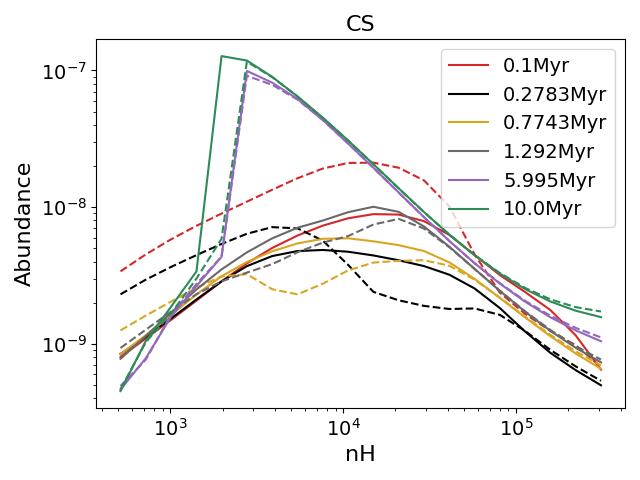}
\caption{Predicted abundances for
CO (top) and CS (bottom) of the 
pre-phase model (solid lines) and
the atomic model (dashed lines) over
the density grid considered in
this work. Different chemical ages
are colour-coded as shown in the
legend.}
\label{fig:atomic_vs_prephase_grid}
\end{figure}

\begin{figure*}[h!]
\centering
\includegraphics[width = .3\textwidth]{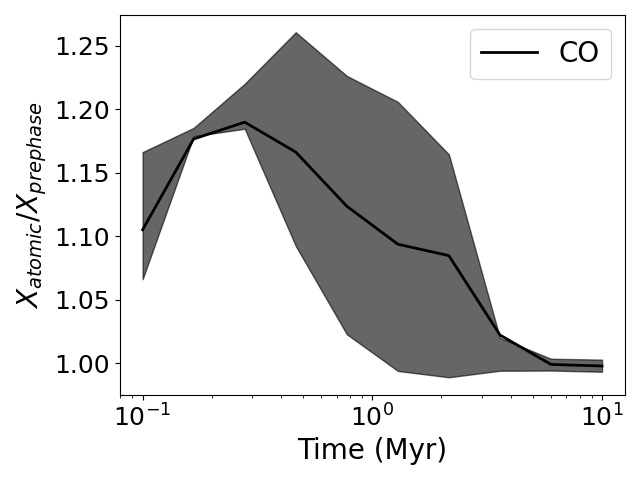}\includegraphics[width = .3\textwidth]{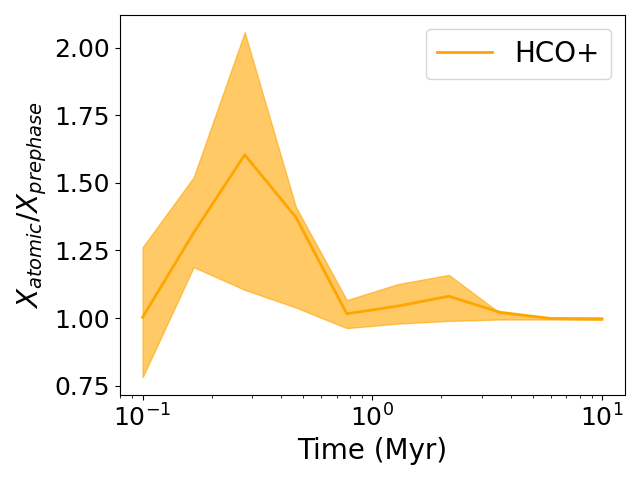}\includegraphics[width = .3\textwidth]{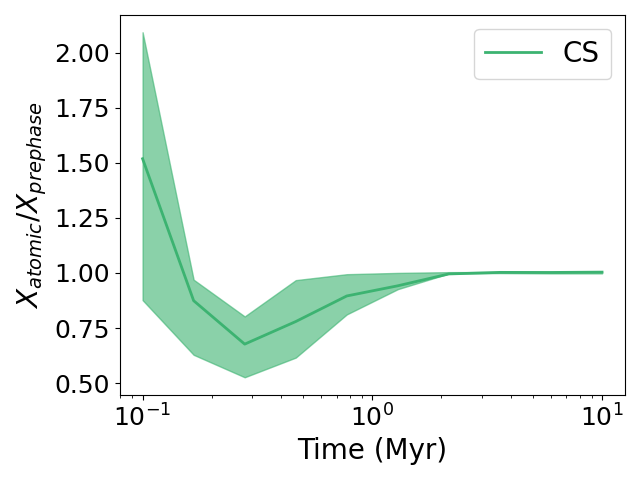}
\includegraphics[width = .3\textwidth]{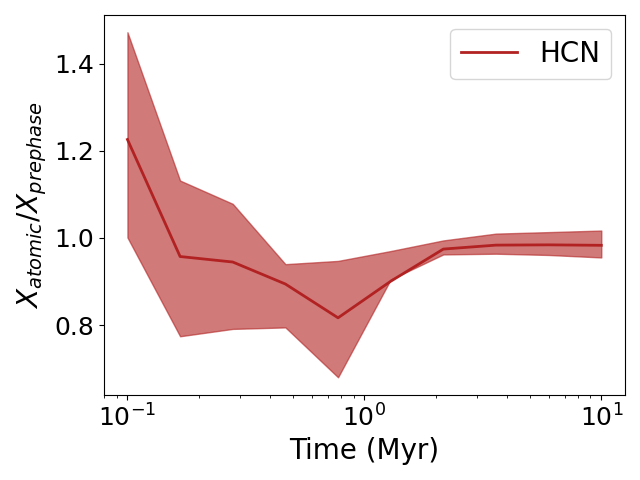}\includegraphics[width = .3\textwidth]{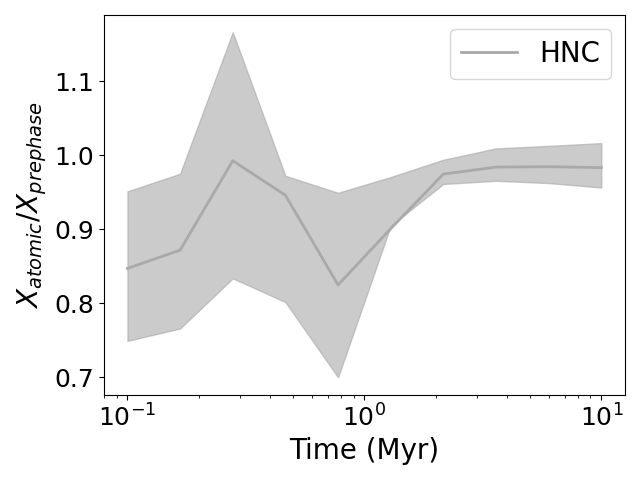}\includegraphics[width = .3\textwidth]{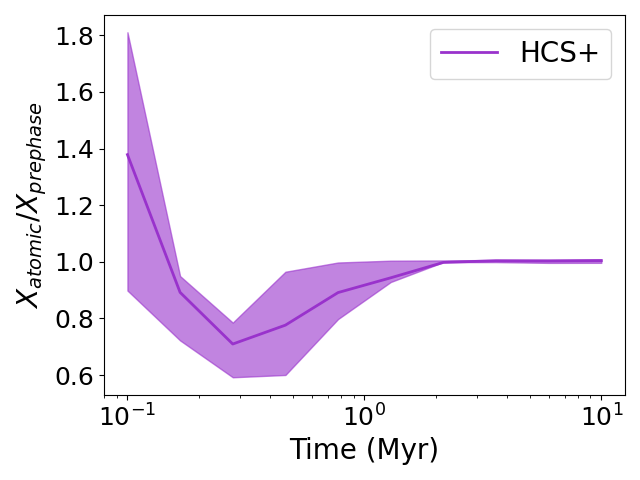}
\includegraphics[width = .3\textwidth]{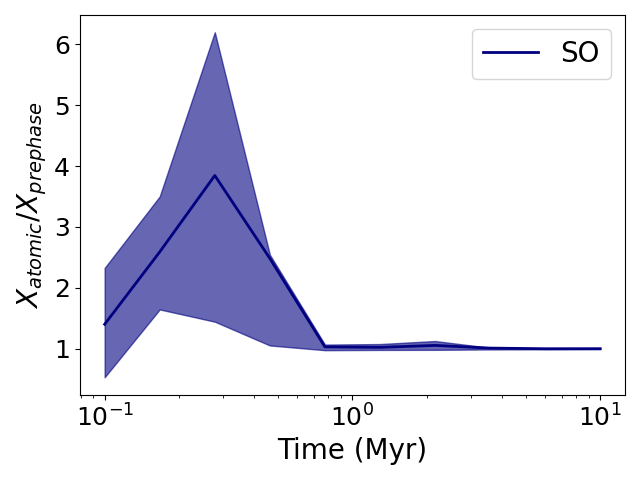}\includegraphics[width = .3\textwidth]{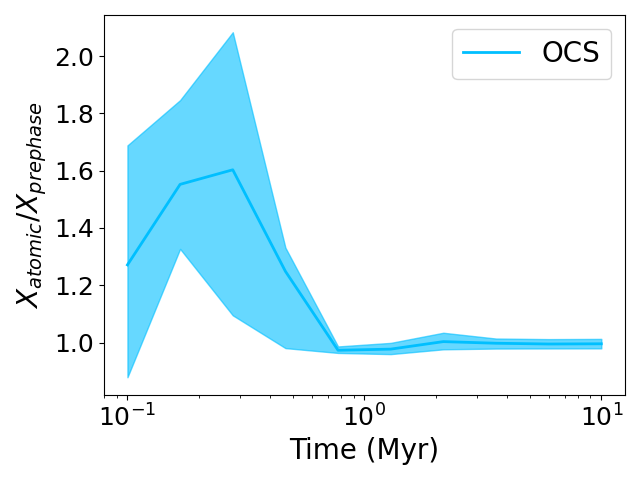}\includegraphics[width = .3\textwidth]{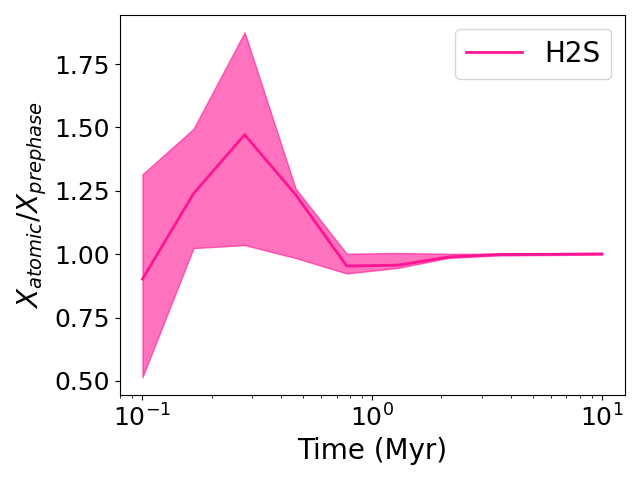}

\caption{Time evolution of the ratio between the abundances
predicted by the model that considers initial atomic
abundances (main paper) and the abundances as given
by the pre-phase model (this section). The solid line
represents the mean ratio, while the shaded area delimits
the interquartile range. For generating these plots,
we only considered density values between percentiles
1.5\% and 98.5\% of the density PDF analysed in the
main core of the paper, which roughly corresponds
to $n_{\rm H} = 1.40 \times 10^{4}$ cm$^{-3}$ and
$n_{\rm H} = 1.20 \times 10^{5}$ cm$^{-3}$.}
\label{fig:relative_abundances}
\end{figure*}

We considered the same density grid as in our model, with 20
density values logarithmically spaced between 
$n_{\rm H} = 10^{2}$ cm$^{-3}$ and 
$ n_{\rm H} =3.084 \times 10^{5}$cm$^{-3}$, and compared the
predicted abundances of the model used in the article,
which we will refer to as the `atomic' model, with those
predicted by the pre-phase model for ages ranging from
0.1 Myr to 10 Myr; the different predictions from these
models for two illustrative cases, CO and CS, are shown
in Fig. \ref{fig:atomic_vs_prephase_grid}.
It is evident for the plots
that the predicted abundances of both
models for CO are quite similar, but there
might be significant differences for the
abundances of sulfur-bearing molecules
such as CS at early ages. In order to quantify
the possible effects of these differences
in abundances between models in the predictions
made in this work, we take the
density values ranging from the 1.5\% and
98.5\% percentiles of our PDF
(shaded region in Fig. \ref{fig:chemical_models}) and analysed
the time evolution of the ratio of 
abundances for our molecules of interest; these predictions
are shown in Fig. \ref{fig:relative_abundances}.

As can be seen in Fig. \ref{fig:relative_abundances}, the
predictions of both models at an age of 1 Myr are almost
equivalent, and differences can arise for chemical ages
between 0.1 Myr and 1 Myr; in the remaining of this section, we will
mainly discuss the differences
at 0.1 Myr and 1 Myr, which are the chemical ages analysed in this
work, but it is important to note that the peak of the
ratio is usually reached at ages around 0.3 Myr.\par

As we have already mentioned, the differences
between models at 1 Myr are barely appreciable, 
but at early ages (0.1 Myr) the predictions might
deviate up to a factor of two, especially
when sulfur-bearing molecules are concerned. As an
illustrative case, we analysed the differences
that might arise for CS, which is one of the molecules
mostly affected by the initial conditions
at 0.1 Myr. If we start
from a medium where all the C is in C$^{+}$
(atomic model), at 0.1 Myr the predicted
abundances might be twice larger than
those given by the pre-phase model, especially
at lower densities (see Fig. \ref{fig:atomic_vs_prephase_grid}). In order to
assess the extent of the influence of these abundance variations in the integrated intensities, we built a molecular line emission
map for CS \trans{2}{1} with the pre-phase model
and the line of sight perpendicular to the
magnetic field, and compared the integrated
line intensities with those of the atomic model.
We observed that the integrated line 
intensities for the atomic model are 
$\sim 1.1 - 1.3$ times larger than those
given by the pre-phase model. However, if we compare
the intensity ratios for the pre-phase model
between the turbulent and uniform models, we
obtain a mean value of 
$I_{\rm TURB} / I_{\rm UNIF} = 0.95$ with an
interquartile range of 0.12, which is
compatible with the results shown in Fig. \ref{fig:relative_intensities_chemistry}.
In consequence, although the chemistry at early
ages is influenced by the initial conditions,
especially for sulfur-bearing molecules, the results
derived in this paper are self-consistent and not
affected by the chemical initial conditions, even
at early ages (0.1 Myr).

\section{Position Velocity diagrams}\label{appendix:PV_diagrams}
\begin{figure*}[h!]
\centering
\includegraphics[width = .25\textwidth]{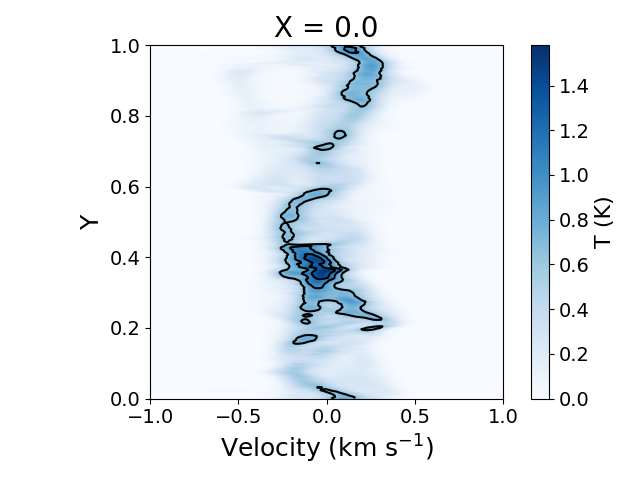}\includegraphics[width = .25\textwidth]{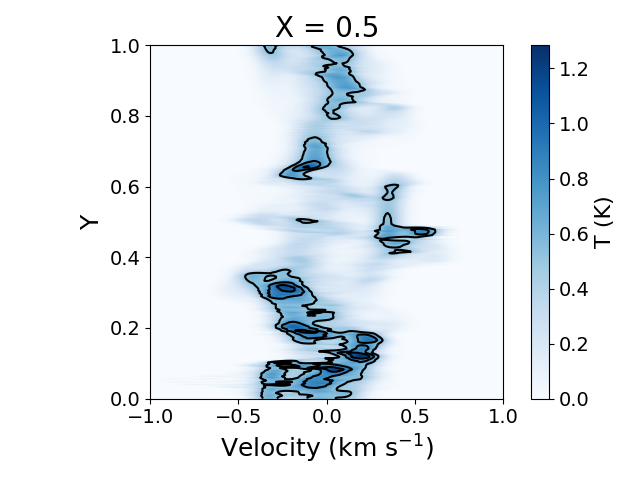}\includegraphics[width = .25\textwidth]{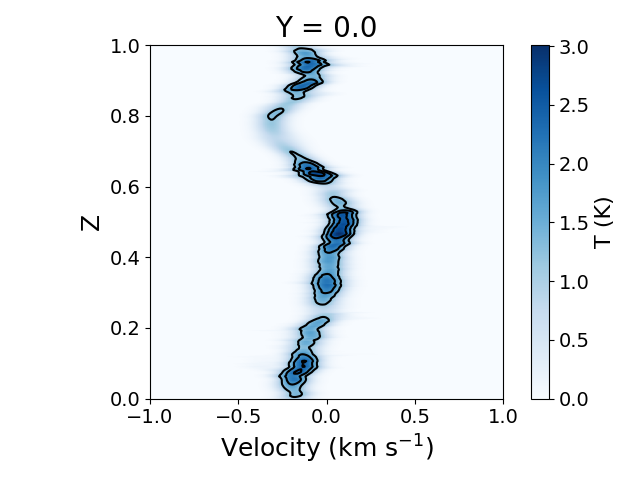}\includegraphics[width = .25\textwidth]{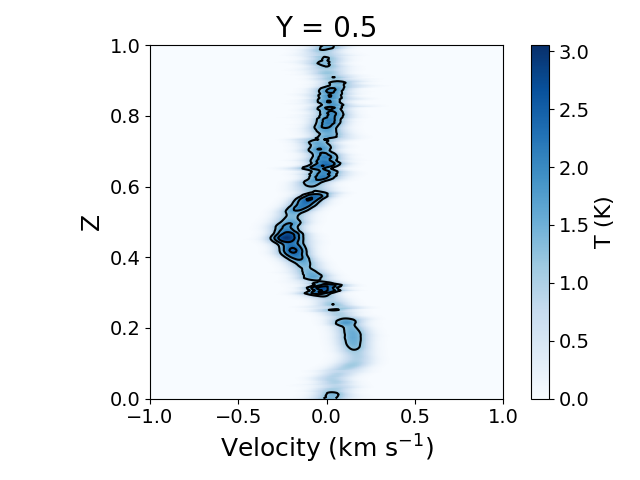}
\caption{Position Velocity diagrams for the line of sight
perpendicular to the magnetic field direction (columns 1 and 2)
and parallel (columns 3 and 4)
in two slices of the cube. The contours
have been drawn at 0.4$T_{\rm max}$, 0.6$T_{\rm max}$ and 0.8$T_{\rm max}$,
where $T_{\rm max}$ is the maximum intensity on each map.}
\label{fig:pv}
\end{figure*}

In this section, supplementary material for the discussion on the effects 
of turbulence on the kinematics (Sec. \ref{sec:line_profiles})
is provided in the form of Position Velocity (PV) diagrams. In Fig. \ref{fig:pv}
we show several PV diagrams
for the two
lines of sight discussed in \ref{sec:line_profiles}, one perpendicular
and one parallel to the magnetic field (left and right panels in
Fig. \ref{fig:column_density_maps}); we decided to also show two different
vertical cuts of each image to highlight the variability of the velocity
structures. Cuts corresponding to the same  
line of sight show a 
similar behaviour: structures in the lines 
of sight perpendicular to the magnetic field are more extended, while
those with a parallel alignment have less velocity dispersion, resulting
in a single filament-like morphology.

\section{Line radiative transfer of molecules with low excitation temperatures}\label{appendix:RADMC3D_bkg}
\begin{figure*}
\centering
\includegraphics[width = \textwidth]{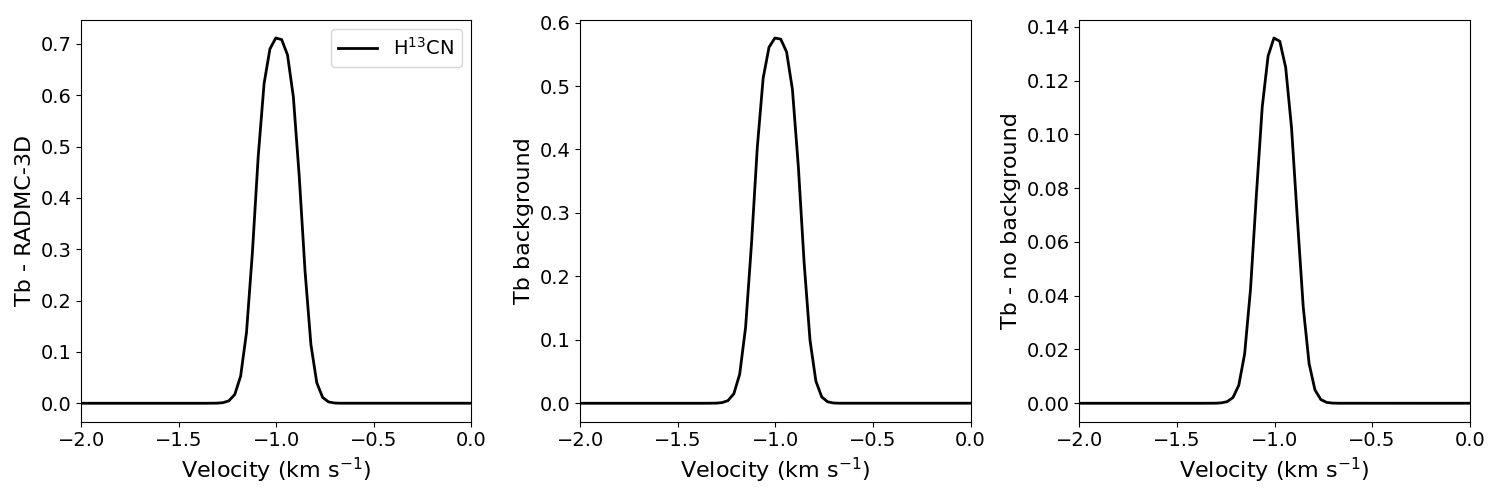}
\caption{Comparison between line profiles. Left: line profile for H$^{13}$CN \trans{1}{0} as given by RADMC-3D.
Middle: line profile for the same setup than left panel, but setting the hydrogen density near to zero, and therefore
corresponding to the background contribution. Right: resulting line profile when we substract the background
(middle plot) to the RADMC-3D profile (left plot). Line profiles like the one on the right are those that we
used for the analysis of  H$^{13}$CN \trans{1}{0} and  H$^{13}$CO$^{+}$ \trans{1}{0}.}
\label{fig:background}
\end{figure*}

As explained in the RADMC-3D manual, the line-integrated mean intensity
for the LVG + EscProb method is given by:

\begin{equation}
J_{i,j} = (1 - \beta _{i,j}) S_{i,j} + \beta _{i,j}J_{i,j}^{\rm bg}
\label{eq:intensity}
\end{equation}

where $J_{i,j}^{\rm bg}$ is the contribution of the background radiation
field to the intensity of the line, which by default is set to
correspond to a blackbody at 2.73 K. Therefore, any line profiles
predicted with this method will include the mean intensity
of this background. In observational works, the background is usually
removed via the on-off technique, that is, observing the source and a nearby position
without emission -- the background, or via frequency switching. For radiative transitions with high excitation temperatures
such as $^{13}$CO \trans{1}{0} the contribution of the background to
the total intensity of the line will be very small -- it is roughly 0.0005 \%.
However, some rotational transitions considered in this study have
low excitation temperatures very similar to the background temperature, and
for those cases the contribution will affect considerably the total
integrated intensity; that is the case for H$^{13}$CN \trans{1}{0} and,
to a lower extent, H$^{13}$CO$^{+}$. The question is, then, how can we
remove this background emission. In the view of equation \ref{eq:intensity},
it is clear that we can recover the radiative emission from the background
making the source function:

\begin{equation}
S_{i,j} = \frac{n_{i}A_{i,j}}{n_{j}B_{j,i} - n_{i}B_{i,j}} = 0
\end{equation}

where $n_{i}$ is the density fraction of the molecule in level $i$ and
$A_{i,j}$, $B_{j,i}$, and $B_{i,j}$ are the usual Einstein coefficients.
We can impose $S_{i,j} \sim 0$ forcing the full molecular fraction to be
at the lowest energy level, that is, supressing the collisional excitation.
Therefore, we run a set of simulations for 
H$^{13}$CN and H$^{13}$CO$^{+}$ where we set the hydrogen densities to
values very near to 0 (actually $n_{H_{2}} \sim 1$ cm$^{-3}$ to avoid
numerical errors). In Fig. \ref{fig:background} we show the differences
that might arise between the line profile of H$^{13}$CO$^{+}$ \trans{1}{0}
as given by RADMC-3D, with the background emission, and when we substract this
background. For our analysis
of  H$^{13}$CN \trans{1}{0} and  H$^{13}$CO$^{+}$ \trans{1}{0},
 we used these substracted profiles.

\section{Additional figure}\label{fig5_additional}
\begin{figure}
\centering
\includegraphics[width = \columnwidth]{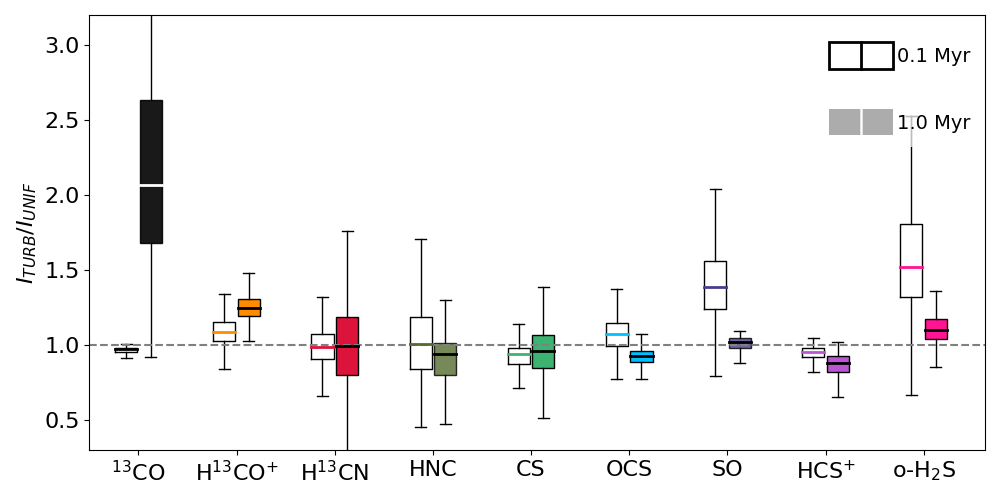}
\caption{Box plot for the ratio of integrated intensity maps 
$I_{\rm TURB} / I_{\rm UNIF}$ for the case of the magnetic field parallel to the
line of sight, where $I_{\rm TURB}$ is the
integrated intensity map for the full post-processed simulation
and $I_{\rm UNIF}$ is the intensity map assuming a constant
hydrogen density (and therefore constant
molecular abundance) along the line of sight.
For each
molecule we have two box plots: the left one corresponds to
the chemical network at
0.1 Myr, and the right one at 1 Myr. 
The y-axis has been scaled so that the main box
ranging from quartiles $Q_{1}$ (25 \%) to $Q_{3}$ (75 \%)
 is always shown and differences
among all molecules can be appreciated, although
in some cases the whiskers extend outside the plot.}
\label{fig:relative_intensities_chemistry_appendix}
\end{figure}

In this appendix, we present the complementary plot to Fig. \ref{fig:relative_intensities_chemistry},
that is, the box plot for the ratio of integrated intensity maps for the case
of the magnetic field parallel to the line of sight (Fig. \ref{fig:relative_intensities_chemistry_appendix}).

\end{appendix}

\end{document}